\documentclass[review,3p,times]{elsarticle}
\usepackage{amsmath}
\usepackage{amssymb}
\usepackage{flafter} 
\usepackage{siunitx}
\usepackage{epstopdf}
\usepackage{caption}
\usepackage{subcaption}
\usepackage{soul}
\usepackage{natbib}
\usepackage{advdate}
\usepackage{float} 
\usepackage[inkscapeformat=png]{svg}
\setcitestyle{authoryear,round,aysep={}}

\graphicspath{ {images/} }

\newlength{\subcolumnwidth}
\newenvironment{subcolumns}[1][0.5\columnwidth]
 {\valign\bgroup\hsize=#1\setlength{\subcolumnwidth}{\hsize}\vfil##\vfil\cr}
 {\crcr\egroup}
\newcommand{\nextsubcolumn}[1][]{%
  \cr\noalign{\hfill}
  \if\relax\detokenize{#1}\relax\else\hsize=#1\setlength{\subcolumnwidth}{\hsize}\fi}

\journal{Journal of Visualization}

\begin{document}
\begin{frontmatter}
\title{Visualizing Three-Dimensional Effects of Synthetic Jet Flow Control}

\author{Adnan Machado\textsuperscript{a}\footnote{Corresponding author \newline \emph{Email:} adnan.machado@mail.utoronto.ca}}
\author{Kecheng Xu\textsuperscript{b}}
\author{Pierre E. Sullivan\textsuperscript{a}}
\address[1]{Mechanical and Industrial Engineering, University of Toronto, 5 King's College Rd, Toronto, Ontario, Canada}
\address[2]{University of Toronto Institute for Aerospace Studies, 4925 Dufferin St, Toronto, Ontario, Canada}
\begin{abstract}
This study investigates the three-dimensionality of synthetic jet flow control over a NACA 0025 profile wing using horizontal and vertical smoke wire visualization. The stalled flow in the baseline case is visualized, providing insights into the shear layer roll-up process, the transition to turbulence, and vortex shedding in the wake. In the controlled flow study, two actuation frequencies, $F^+=1.18$ and $F^+=11.76$, are investigated, with a focus on spanwise control authority and the role of coherent structures in flow reattachment. The results indicate that while the control is effective at the midspan over the entire chord length, its effect diminishes with increasing distance from the midspan. Both control cases result in significant spanwise velocities, observed by a contraction of the flow towards midspan. Lastly, the high-frequency actuation results in unique small-scale structures at the shear layer-freestream interface.
\newline
\newline
\emph{Keywords:} Active flow control, Synthetic jet, Smoke wire visualization
\end{abstract}
\end{frontmatter}

\noindent
\textbf{Competing Interests:} The authors declare that they do not have competing interests. \newline
\textbf{Funding:} This research was funded by Natural Sciences and Engineering Research Council of Canada (NSERC) grant number RGPIN-2022-03071 and the Digital Research Alliance of Canada (4752)

\section{Introduction}
\label{sec:intro}
Flow control of reattaching separated flows has many applications, such as preventing stall in low-speed, high-altitude aircraft for surveying and stealth, and improving the aerodynamics of wind turbine blades with short chord lengths. Synthetic jet actuators (SJAs) are zero-net-mass-flux devices that add momentum to the flow via periodic suction and blowing. In contrast to continuous jets, SJAs do not require plumbing and reservoirs which makes them ideal for flow control. Furthermore, \citet{DiCicca2007,Smith2003, Seifert1999} demonstrated that synthetic jets outperform continuous jets in fluid entrainment and mixing. Numerous studies have demonstrated the ability of synthetic jets to reattach flows on stalled airfoils, resulting in significant drag reduction and lift increase \citep{Feero2015,Rice2018,Andino2015,Greenblatt2000,Amitay2001,Glezer2005,Yang2022,Xu2023}. A widely adopted design, rectangular slot-style SJAs, require significant modification to the wing which could cause structural issues. An alternate design, employing MEMS-based microblowers with small, circular nozzles, could minimize the machining required on wings. These SJAs can be installed with minimal impact on the airfoil surface, resulting in a more aircraft-ready design.

The process of flow reattachment is primarily attributed to the presence of spanwise vortical structures that convect over the suction side of the airfoil. These structures entrain high-momentum fluid from the freestream into the separated shear layer \citep{Greenblatt2000, Rice2018, Xu2023, Salunkhe2016}. This transported momentum energizes the shear layer, enabling the flow to overcome the adverse pressure gradient, leading to flow reattachment. Clearly, the understanding of the formation, evolution, and dissipation of these coherent structures is critical in designing effective active flow control systems. The formation of these structures is not well agreed upon and may be dependent on both flow conditions and SJA parameters. A recent numerical study by \citet{Tousi2021} showed that spanwise vortices are formed directly by the oscillatory motion of the synthetic jets, which merge to create larger vortices. Conversely, \citet{Salunkhe2016} and \citet{Tang2014} concluded that spanwise vortices are created by the separated shear layer, but are induced by actuation. \citet{Toyoda2009} states that mixing happens in two stages, 1) large-scale entrainment of the surrounding fluid, and 2) small-scale turbulent mixing via vortex breakdown. The latter is consistent with the turbulent boundary layers observed by \citet{Salunkhe2016}, \citet{Yang2022}, and \citet{Xu2023} in the reattached flows.

The actuation frequency of the SJA is an important parameter for flow control as it dictates the rate of formation, size, and spacing of coherent spanwise vortical structures. The actuation frequency is non-dimensionalized as $F^+=fc/U_\infty$, where $f$ is the forcing frequency, $c$ is the chord length, and $U_\infty$ is the freestream velocity. Prior studies have exploited the natural frequencies of the baseline flow via actuation at the wake shedding frequency $(\mathrm{St_w}\approx \mathcal{O}(1))$ or the shear layer roll-up frequency $(\mathrm{St_{sl}}\approx \mathcal{O}(10))$ \citep{Amitay2002,Glezer2005,Yarusevych2008,Feero2015,Yang2022,Xu2023}. These studies show that actuation in the order of the wake shedding frequency results in large-scale spanwise vortices. This results in time-averaged lift recovery, but is associated with unsteady flow dynamics, which is not desirable in aviation. In contrast, actuation in the order of the shear layer instability creates many smaller vortices, resulting in quasi-steady flow reattachment \citep{Glezer2005,Xu2023}. \citet{Glezer2005} demonstrated that actuation at $F^+\approx \mathcal{O}(1)$ results in a coupling of the shear layer instability and the natural wake instability. Furthermore, if the actuation frequency is sufficiently high, the natural instabilities of the airfoil are decoupled from the actuation, and the flow can be completely reattached in the absence of phase-coherent structures.

Another important parameter of SJAs is the blowing strength, which is often represented by a non-dimensional blowing ratio, $C_B=\overline{U_j}/U_\infty$, where $\overline{U_j}$ is the time-averaged jet velocity. As  well, momentum coefficient can be used~\citep{Amitay2001}, \begin{equation}
    C_\mu=\frac{\overline{I}_j}{\frac{1}{2}\rho_oA_fU_\infty^2}
\end{equation}
where $\overline{I_j}$ is the time-averaged jet momentum, and $A_f$ is the projected control area.
\citet{Yang2022} demonstrated that increasing the blowing ratio delayed the dissipation of the spanwise vortices, and forced them on a downward trajectory, resulting in better control. The control is said to be saturated when a further increase to the blowing magnitude does not result in aerodynamic improvements, however, \citet{Feero2017a} showed that while the control effects may be saturated at midspan, increasing the blowing ratio further may improve the spanwise extent of the control. This highlights the need for three-dimensional analysis of the controlled flow across the span of the wing. Additionally, parameters such as jet location and angle influence the interaction between the synthetic jet and the shear layer. The results of \citet{Feero2017b} showed that the SJAs should be placed as upstream as possible, while \citet{Tang2014} observed that actuation close to the separation point was most effective. \citet{Amitay2001} discovered an interaction between the optimal jet angle, and the angle of attack, suggesting that the optimal actuation parameters depend on the state of the separated flow.

Considering the large parameter space and the interaction of these parameters with varying flow conditions, optimizing SJAs for flow control becomes a daunting task. The majority of experimental research focuses on the effects of flow control at the symmetry plane of the wing, employing quantitative techniques such as particle image velocimetry, hot-wire anemometry, and surface pressure measurement. However, these techniques seldom assess the spanwise control authority, leaving a gap in the understanding of the three-dimensional controlled flow.

In this study, the smoke wire technique is employed to visualize the baseline and controlled flow around the airfoil, with the goal of uncovering the nature of the three-dimensional flow and identifying the effects of coherent structures. Multi-plane smoke wire imaging is used to visualize the controlled flow at various spanwise positions. Moreover, a horizontal smoke wire is utilized to visualize flow patterns involving spanwise velocity components, while also providing insights into the spanwise-uniformity of the flow. This comprehensive flow visualization approach enables the observation of both small-scale features, such as shear layer instabilities, as well as large-scale structures in the wake, capturing their interactions and evolution. The baseline flow is characterized, followed by an analysis of the controlled flow field at both low- and high-frequency actuation. 

\section{Experimental Method}
\label{sec:expmeth}
\subsection{Wind Tunnel}
Experiments are conducted in the low-turbulence recirculating wind tunnel at the Department of Mechanical and Industrial Engineering, at the University of Toronto. The test section has dimensions of 5 m $\times$ 0.91 m $\times$ 1.22 m and features acrylic windows on the top and side walls for observation and measurement. The flow passes through seven screens and a 12:1 contraction before entering the test section. The wind tunnel is capable of producing speeds between 3--18 m/s with a turbulence intensity of 1\%. For the experiments conducted, the wind tunnel was operated at a freestream velocity of $U_\infty=5.1$ m/s, which corresponds to a chord-based Reynolds number of \(\mathrm{Re}_c=10^5\).

\subsection{Airfoil model}
A NACA 0025 profile wing, designed by \citet{Feero2015}, is situated in the test area of the wind tunnel with the leading edge approximately 40 cm from the test section inlet (Figure \ref{fig:smokewirelocations}). The aluminum wing has an aspect ratio of approximately 3, with a wingspan of $b=885$ mm, and a chord length of $c=300$ mm. The wing spans the entire width of the test section and features circular end plates which isolate it from the boundary layer at the wind tunnel walls, as shown by \citet{Feero2017b}. The wing is made up of 3 parts, with a hollow center third to house the sensors and actuators. In the center piece, there is a 317 mm $\times$ 58 mm rectangular cutout where the microblower array is installed, with a flush 0.8 mm hole for the nozzle of each SJA. The angle of attack is set to $\alpha=10^\circ$, such that the flow separates at approximately 12\% chord with the specified flow parameters.

\subsection{Microblowers}
\begin{figure}[h]
   \label{fig:expsetup}
\begin{subcolumns}[0.5\textwidth]
  \begin{subfigure}{.5\textwidth}
  \includegraphics[width=\linewidth]{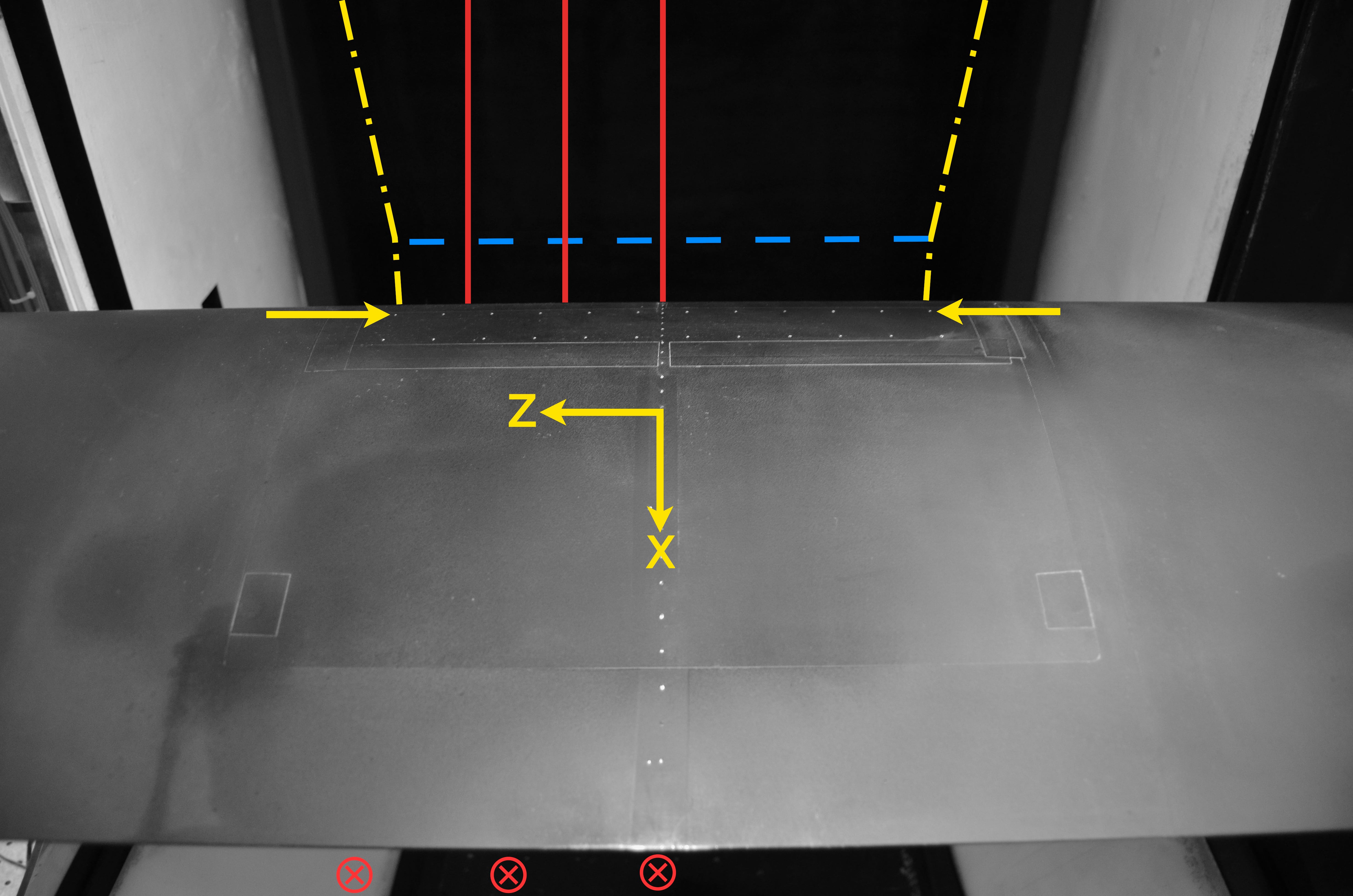}
  \caption{}
  \label{fig:smokewirelocations}
 \end{subfigure}
\nextsubcolumn
\hfill
\begin{subfigure}{0.455\textwidth}
\includegraphics[width=\linewidth]{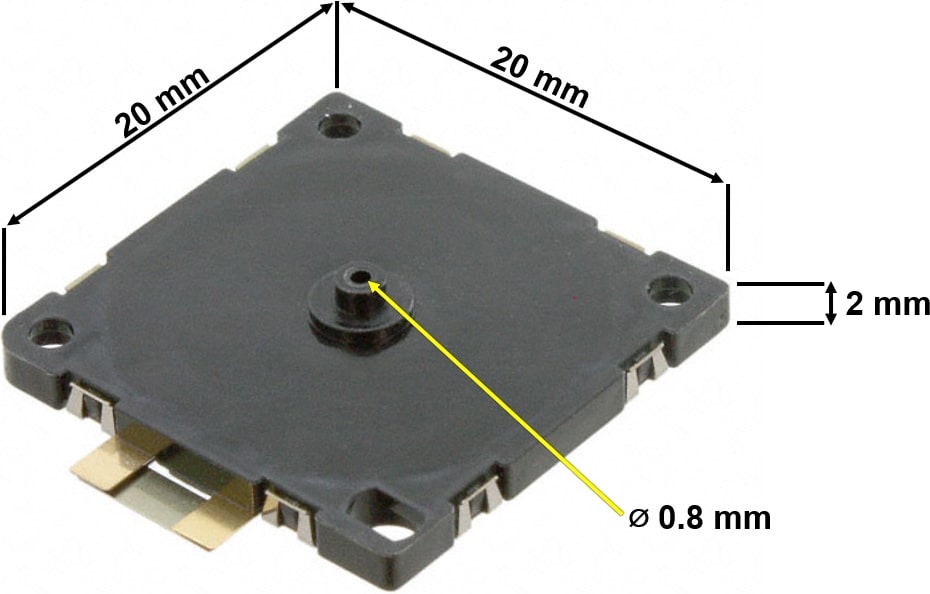}
\centering
\caption{}
\label{fig:SJA}
\end{subfigure}
  \end{subcolumns}
    \begin{subfigure}{\textwidth}
\includegraphics[width=0.7\linewidth]{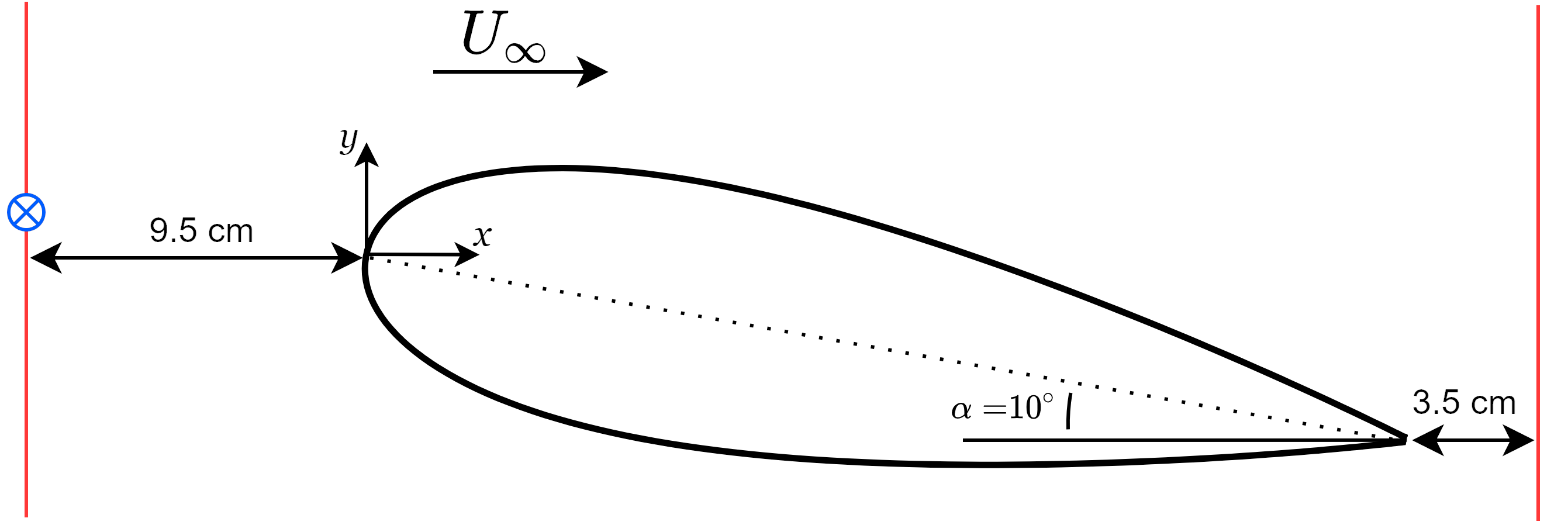}
\centering
\caption{}
\label{fig:depiction}
    \end{subfigure}
     \caption{a) The airfoil in the test section with the locations of smoke wires, b) the Murata microblower, c) depiction of the side view of the airfoil and smoke wire setup}
\end{figure}

The SJAs used are the commercially available Murata MZB1001T02 microblowers, pictured in Figure \ref{fig:SJA}, and are embedded underneath the surface of the airfoil. The array consists of two rows of 12 SJAs located at 10\% and 18\% chord. However only the upstream row is used in this experiment, as indicated by the arrows in Figure \ref{fig:smokewirelocations}. The SJA operates between 5--30 V and has a drive resonant frequency between 24--27 kHz. \citet{Xu2023} observed that the mean centerline velocity of the synthetic jet reaches a maximum when driven at a frequency of 25.1 kHz. However, due to the velocity response being right-skewed, the carrier frequency is chosen as $f_c=25.5$ kHz to ensure a stable jet velocity. In this investigation, the SJAs are burst modulated at two excitation frequencies $f_e=20$ Hz and 200 Hz, corresponding to non-dimensional frequencies of $F^+=1.18$ and 11.76, respectively. Square waveforms are used for both the carrier and modulation frequency, with a duty cycle of 50\%. The SJAs are operated at 20 $\mathrm{V}_{pp}$, corresponding to a momentum coefficient of $C_\mu=2.0\times10^{-3}$~\citep{Xu2023}.

\subsection{Smoke Flow Visualization}
Smoke flow visualization was performed with 1) upstream and downstream vertical smoke wires indicated in red in Figure \ref{fig:smokewirelocations}; and 2) a single upstream horizontal smoke wire shown in blue. Vertical smoke wires were installed 9.5 cm upstream of the leading edge and 3.5 cm downstream of the trailing edge of the wing, as illustrated in Figure \ref{fig:depiction}. A pair of upstream and downstream smoke wires could be installed at three spanwise configurations: midspan, 5 cm off-center, or 10 cm off-center ($z/c=[0, 0.17, 0.33]$). The vertical smoke wires were kept under tension by attaching small weights to them underneath the wind tunnel. The horizontal smoke wire was installed upstream of the wing, along the chord line, allowing for visualization of the flow at the edge of the shear layer. The smoke wire length spanned slightly larger than the SJA array. Figure \ref{fig:smokewirelocations} shows how the horizontal wire (blue dashed line) is installed between two vertical support wires (yellow dash-dot lines). The horizontal wire was kept taut by making it slightly shorter than the distance between the vertical support wires so that they bowed inward. This also helped ensure that the horizontal wire remained in the same position across all tests.

To prevent vortex shedding from the wire, the Reynolds number should be kept below $\mathrm{Re}_d=49$ \citep{Batill1981}. Steel wire with a diameter of 0.12 mm was used as it is thin enough to maintain laminar flow at the flow speed used, yet still strong enough to withstand thermal fatigue under constant tension. However, vortex shedding still occurred when the wire was coated with large oil droplets. To prevent this, a large droplet was applied at the top of the wire which quickly falls due to gravity, leaving behind a trail of droplets just slightly larger than the wire. This method allowed for quick vaporization of the oil droplets, creating fine and uniform smoke streams. For the horizontal wire, a cotton swab was used to manually apply oil before each test.

The vertical smoke wire circuit was wired in parallel and had a measured resistance of 118 $\Omega$ per branch, while the horizontal smoke wire circuit had a resistance of 160 $\Omega$. A variable transformer was used to apply 125 V across the smoke wires, corresponding to a heat flux of approximately 260 kW/m$^2$ and 140 kW/m$^2$ for the vertical and horizontal wires, respectively. This heat flux resulted in sufficiently dense smoke creating quality streaklines for approximately 2 seconds. After many heating cycles during experimentation, the measured resistance of the wire was unchanged, indicating that the wire did not degrade over time.

A Nikon SB-800 speedlight was set atop the wind tunnel which provided a short burst of illumination, necessary for capturing high-frequency flow phenomena. The speedlight flash duration was set to 1/5900 seconds. For the vertical smoke wire imaging, the contrast was enhanced by using foil-lined barriers to shape the flash into a thin, long rectangle oriented in the direction of the flow. This allowed for the smoke streams to be fully illuminated while maintaining a dark background. A Nikon D7000 DSLR camera was used to capture images from outside the wind tunnel. For the vertical smoke wire configurations, the camera was positioned on the side of the wind tunnel, and for the horizontal smoke wire configuration, the camera was placed atop the wind tunnel, providing a top-down view of the flow. The photos were taken with an aperture of f/8, and an ISO of 100 and 640 for the vertical and horizontal configurations, respectively. The shutter speed is irrelevant as the exposure is limited by the short duration of the speedlight. With the specified camera and flash settings, the camera could be operated in burst mode at 6 images per second, facilitating the capture of the streaklines at their peak density. The camera and the flash were synchronized using wireless radio transmitters. Lastly, the brightness and contrast of the images were edited to enhance visualization. Due to reflection, the overhead imaging had the least contrast making the smoke streams barely visible, and thus required a further step in editing. Initially, a smoke-free image was captured and later used for image subtraction along with the smoke photos. This process, followed by further contrast adjustment, resulted in clear smoke streams in the region of interest.

\section{Results}
\label{sec:results}
\subsection{Baseline flow}
\begin{figure}[t!]
    \begin{subfigure}{0.51\textwidth}
        \includegraphics[width=\linewidth]{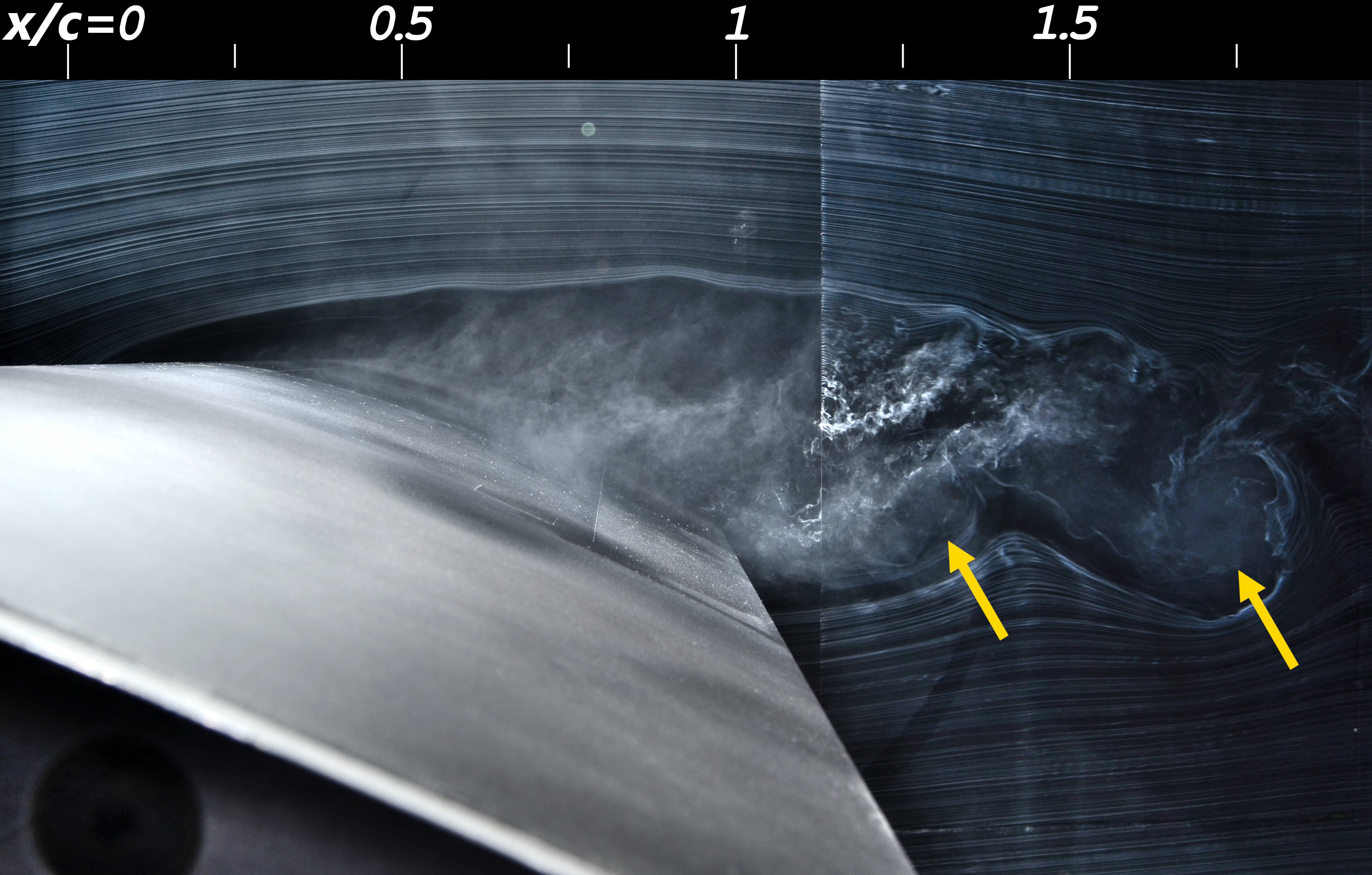}
        \setcounter{subfigure}{0}
        \vspace*{-8mm}
        \caption{}
        \label{fig:baselineA}
    \end{subfigure}
    \begin{subfigure}{0.49\textwidth}
    \includegraphics[width=\linewidth]{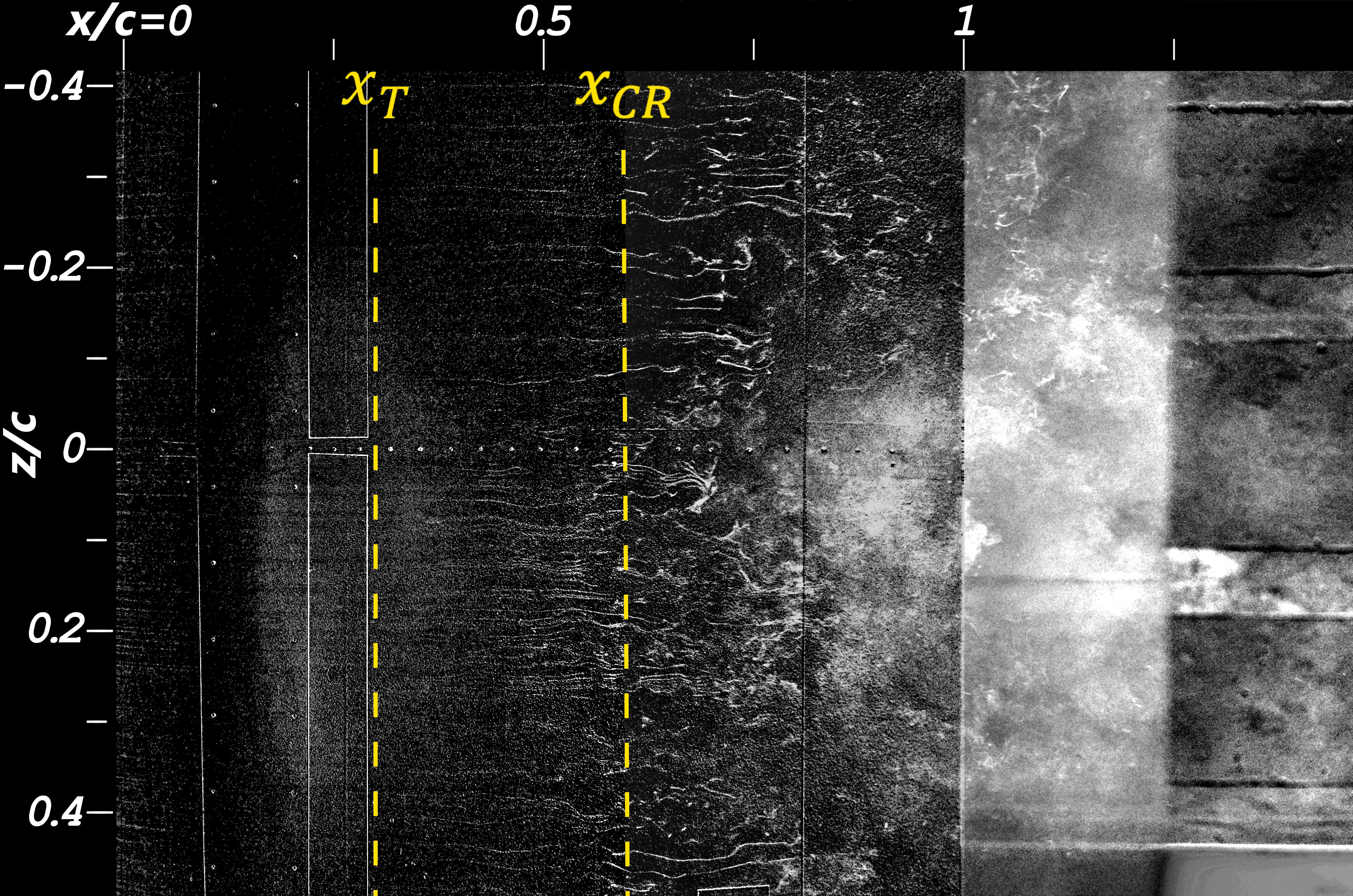}
        \setcounter{subfigure}{2}
        \vspace*{-8mm}
        \caption{}
        \label{fig:baselineC}
    \end{subfigure}
    \begin{subfigure}{0.577\textwidth}
        \includegraphics[width=\linewidth]{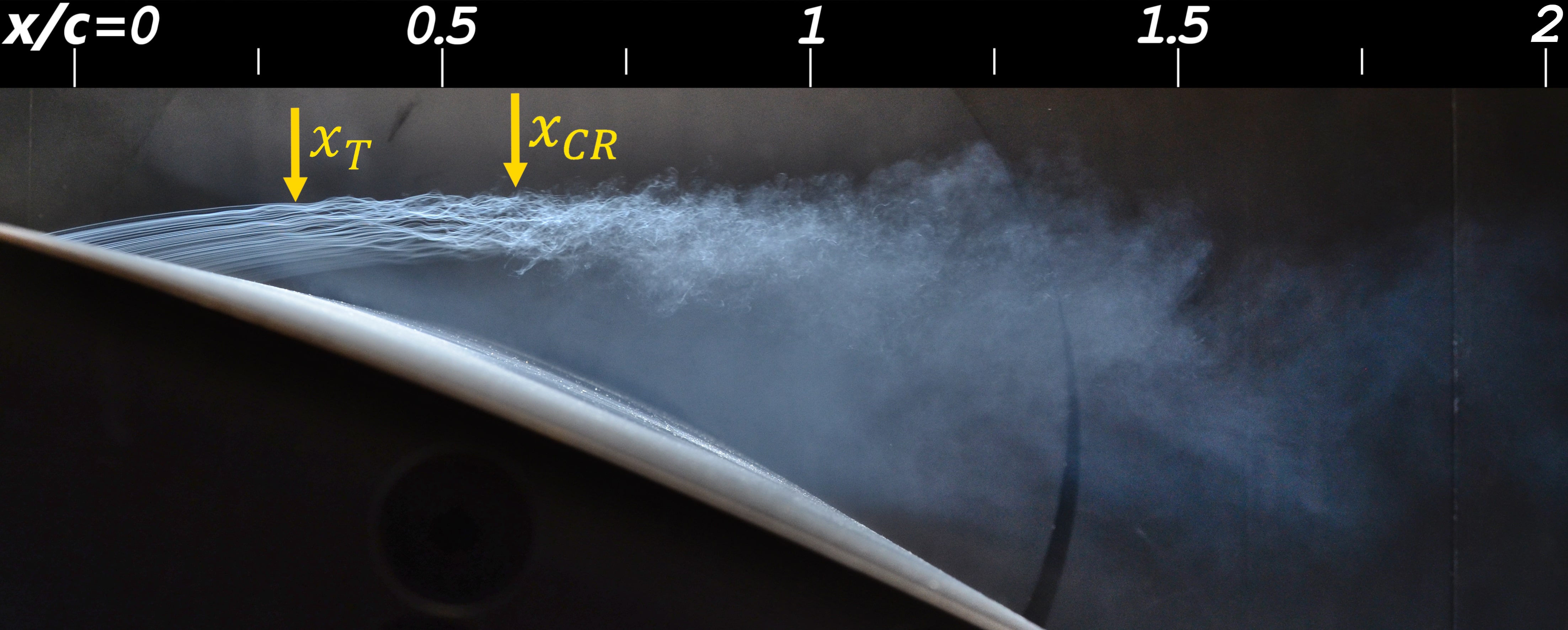}
        \setcounter{subfigure}{1}
        \vspace*{-8mm}
        \caption{}
        \label{fig:baselineB}
    \end{subfigure}
    \begin{subfigure}{0.423\textwidth}
    \includegraphics[width=\linewidth]{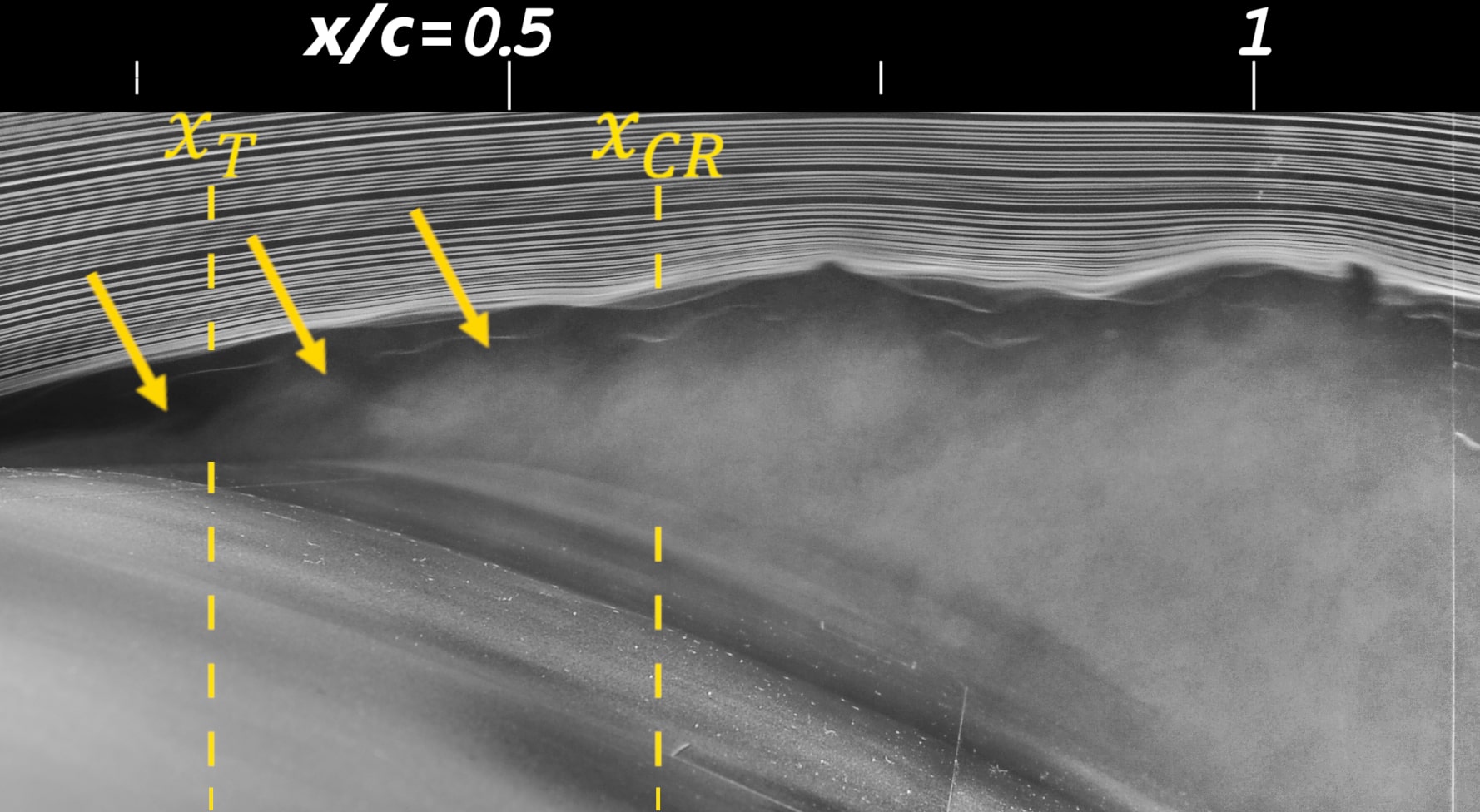}
        \setcounter{subfigure}{3}
        \vspace*{-8mm}
        \caption{}
        \label{fig:baselineD}
    \end{subfigure}
    \caption{Smoke visualization of the baseline flow. a) Upstream and downstream smoke wires, b) side view with horizontal smoke wire, c) overhead view with horizontal smoke wire, d) zoomed in view of the shear layer roll-up}
    \label{fig:baseline}
\end{figure}

The baseline smoke wire image in Figure \ref{fig:baselineA} indicates that the airfoil is stalled, evidenced by the high trajectory of the laminar streaklines. The smoke from the rear smoke wire reveals that the adverse pressure gradient induces flow reversal at the trailing edge, creating a recirculation area extending upstream to the separation point. The diffuse smoke in the recirculation area indicates a highly turbulent shear layer. This low-velocity fluid over the suction side of the airfoil greatly reduces lift. Additionally, large-scale vortex shedding is seen from the trailing edge of the airfoil. Two counter-clockwise vortices are observed in the wake, labelled by the yellow arrows in Figure \ref{fig:baselineA}, and are surrounded by areas of lower strength clockwise vorticity, forming a non-uniform vortex street. The large wake associated with the separated flow results in a high pressure drag. With the same experimental setup, \citet{Xu2023} observed a peak in the velocity spectra of the wake at $\mathrm{St_w}\approx\mathcal{O}(1)$, which can be attributed to the frequency of the observed vortex shedding.

Through a series of flow visualizations, the transition to turbulence in the baseline flow can be investigated in detail. The horizontal smoke wire image of the baseline case (Figure \ref{fig:baselineB}) reveals flow patterns at the shear layer-freestream interface. Slightly downstream of the separation point, at $x/c\approx0.3$, the streaklines begin to display a sinusoidal flow pattern, characteristic of the Kelvin-Helmholtz (KH) instability. The onset of this instability marks the start of the turbulent transition region, $x_T$. The rapid growth of these perturbations results in a fully turbulent flow by $x_{CR}\approx0.6c$, indicated by the diffusion of smoke streams. The overhead view of the horizontal smoke wire (Figure \ref{fig:baselineC}) reveals a predominantly two-dimensional flow in the transition region, as minimal spanwise fluctuations appear in the streaklines. Downstream of $x_{CR}$, the turbulent flow takes on a more 3-dimensional nature, evident by the increased spanwise fluctuations and the eventual dissipation of the streaklines further downstream. In Figure \ref{fig:baselineD}, the recirculated smoke from the rear smoke wire allows for visualization of structures inside the shear layer. It can be seen that the freestream flow sweeps over the KH waves, resulting in breaking waves, three of which can be seen in the figure, and are labelled by the downward yellow arrows. Through this two-dimensional mechanism of shear layer roll-up, spanwise vortices form over the airfoil, and then dissipate into small-scale turbulence. The visualization of the transition to turbulence aligns with the spectra of fluctuating velocity, as measured by \citet{Yarusevych2008}, using the same airfoil with identical flow conditions. The results showed the growth of a broadband spectral peak centered about $\mathrm{St_{sl}}\approx\mathcal{O}(10)$, followed by a turbulent spectral profile downstream at $x/c=0.59$, suggesting the formation, growth, and decay of shear layer vortices. \citet{Balzer2010} compared the evolution of vortex structures on a stalled airfoil generated by 2-D and 3-D DNS and showed that spanwise velocity fluctuations are responsible for the decay of these coherent structures. Lastly, as seen in Figure \ref{fig:baselineC}, the transition to turbulence is observed to be non-uniform across the span, a phenomenon also observed by \citet{Kirk2017} at higher angles of attack. The size of the separated shear layer, and thus the amount of turbulent kinetic energy, increases with $\alpha$. This results in spanwise fluctuations that are sufficiently large to transport streamwise momentum across the span, resulting in a non-uniform spanwise transition to turbulence.
\subsection{Controlled flow}

\begin{figure}[t!]
    \centering
    \begin{subfigure}{0.495\textwidth}
        \includegraphics[width=\linewidth]{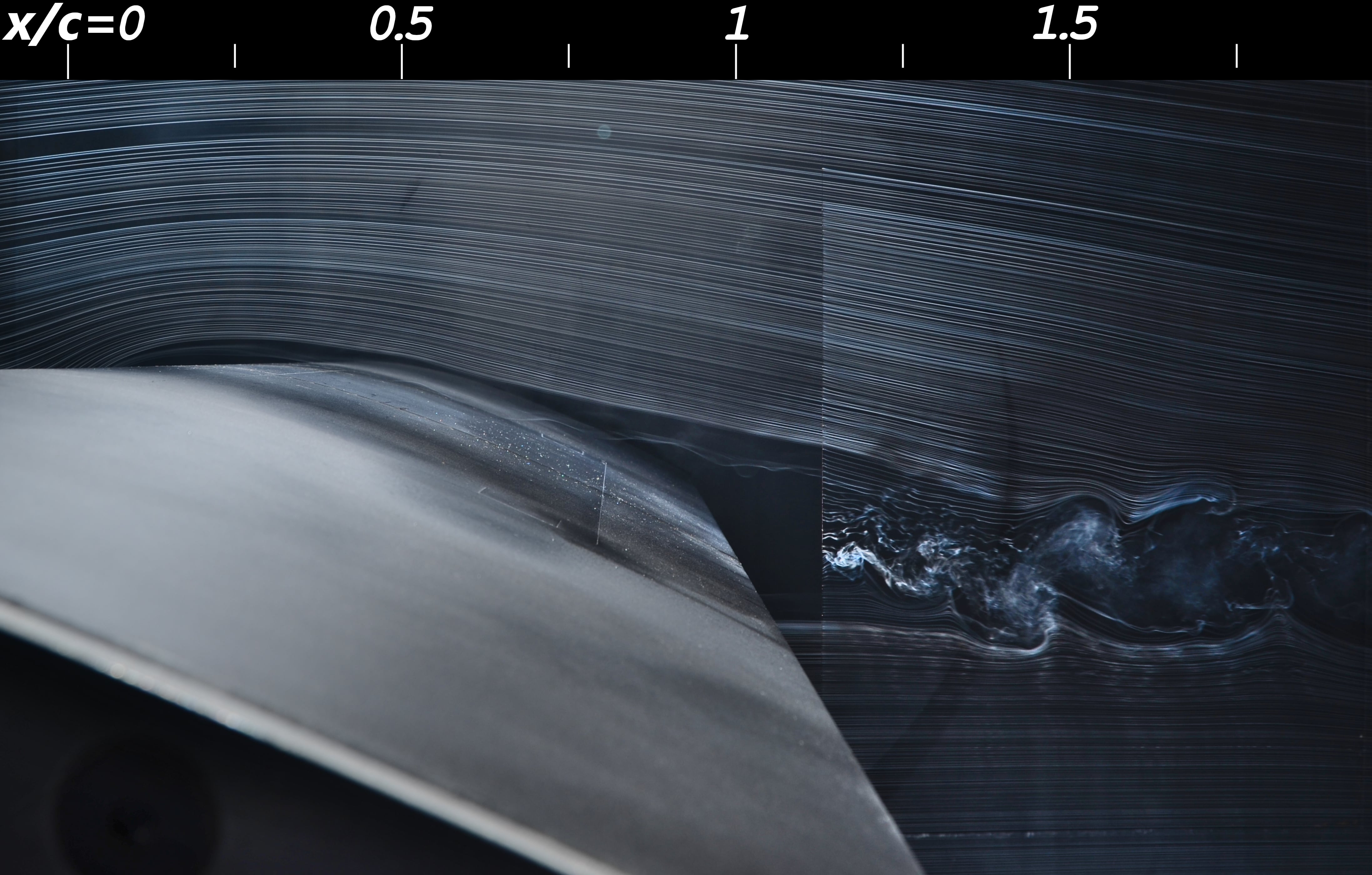}
        \caption{$F^+=1.18$, measured at midspan}
        \label{fig:20Hz0}
    \end{subfigure}
    \begin{subfigure}{0.495\textwidth}
        \includegraphics[width=\linewidth]{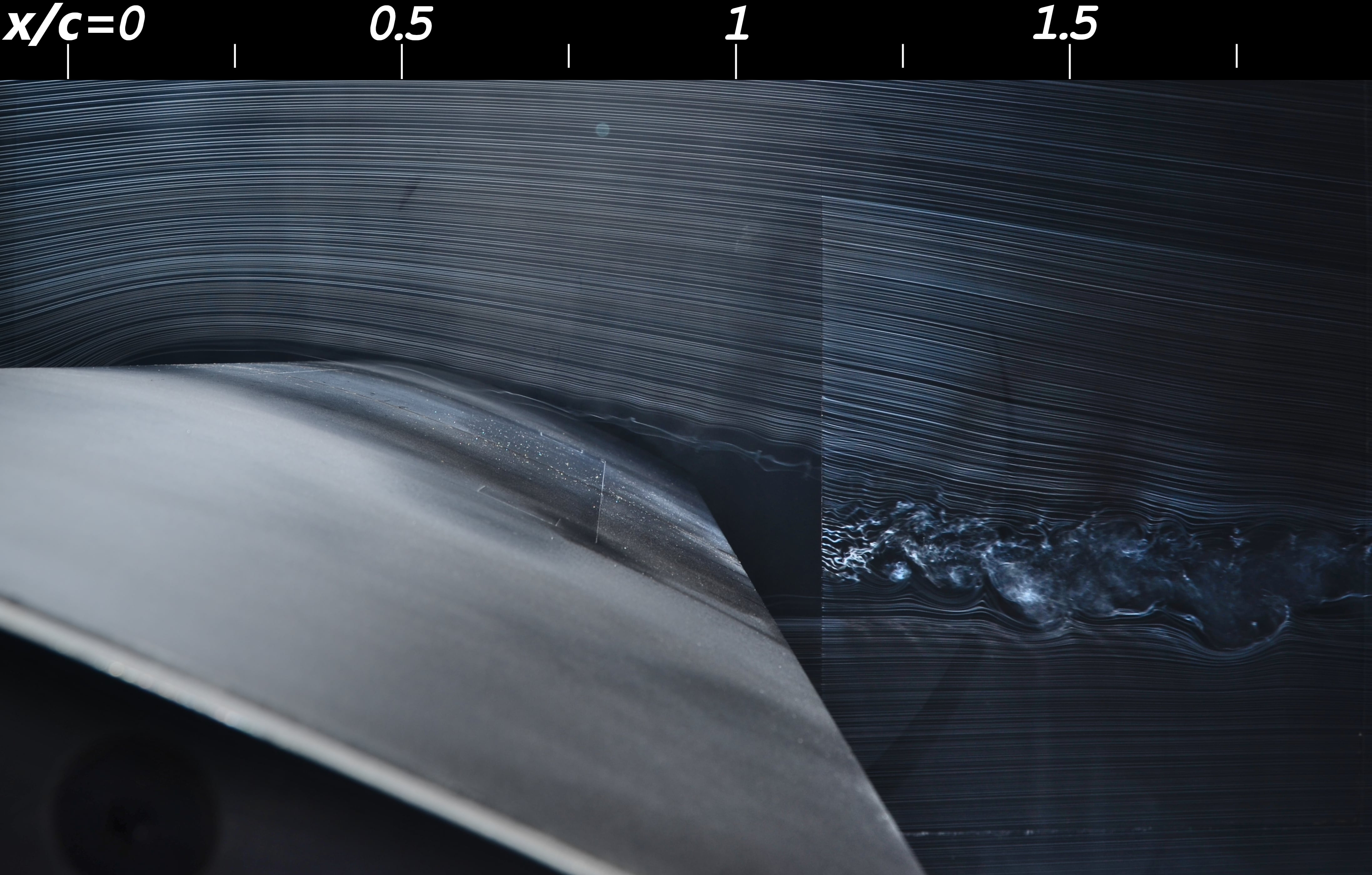}
        \caption{$F^+=11.76$, measured at midspan}
        \label{fig:200Hz0}
    \end{subfigure}
    \begin{subfigure}{0.495\textwidth}
        \includegraphics[width=\linewidth]{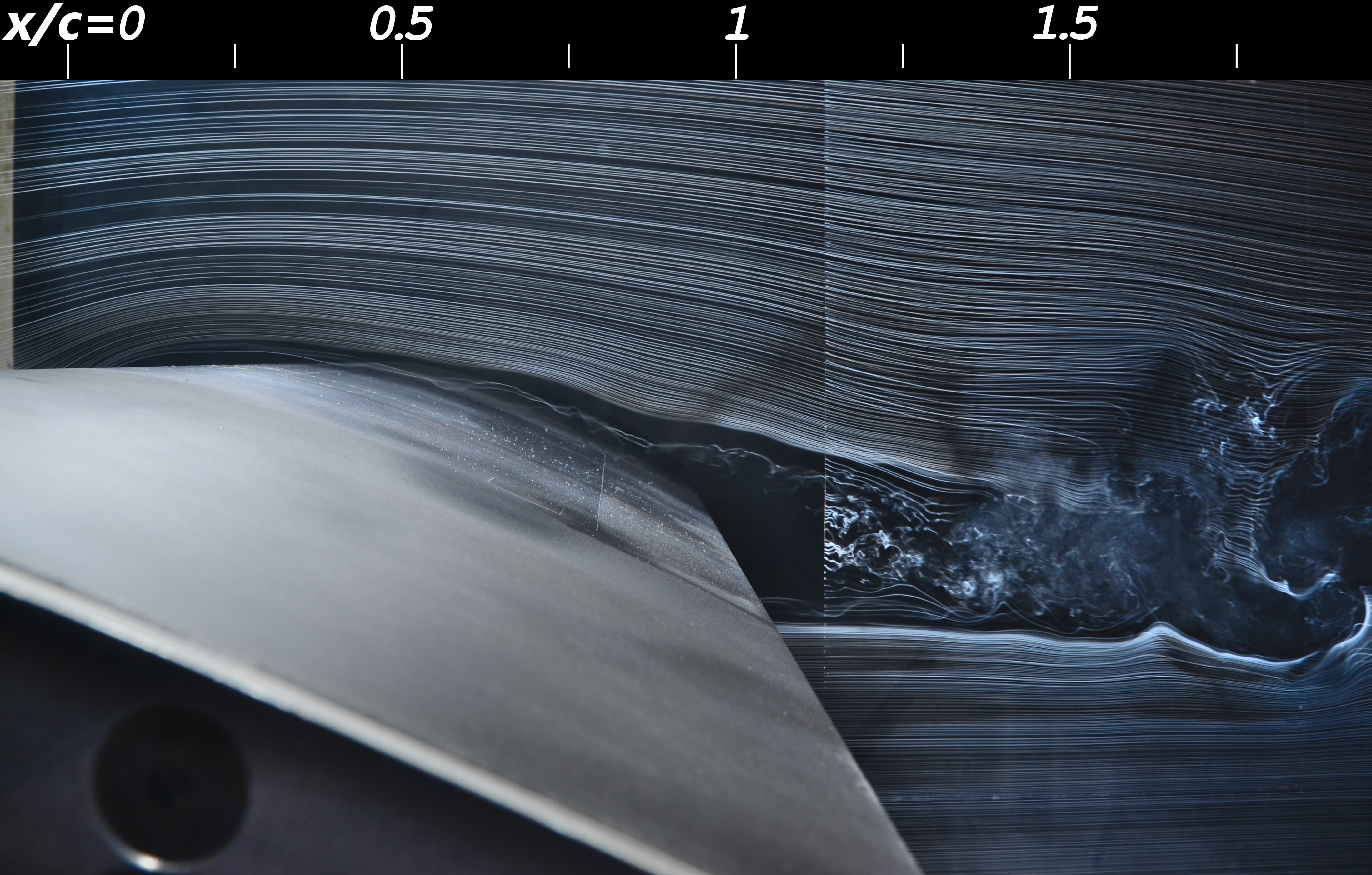}
        \caption{$F^+=1.18$, measured at $z/c=0.17$}
        \label{fig:20Hz5}
    \end{subfigure}
        \begin{subfigure}{0.495\textwidth}
        \includegraphics[width=\linewidth]{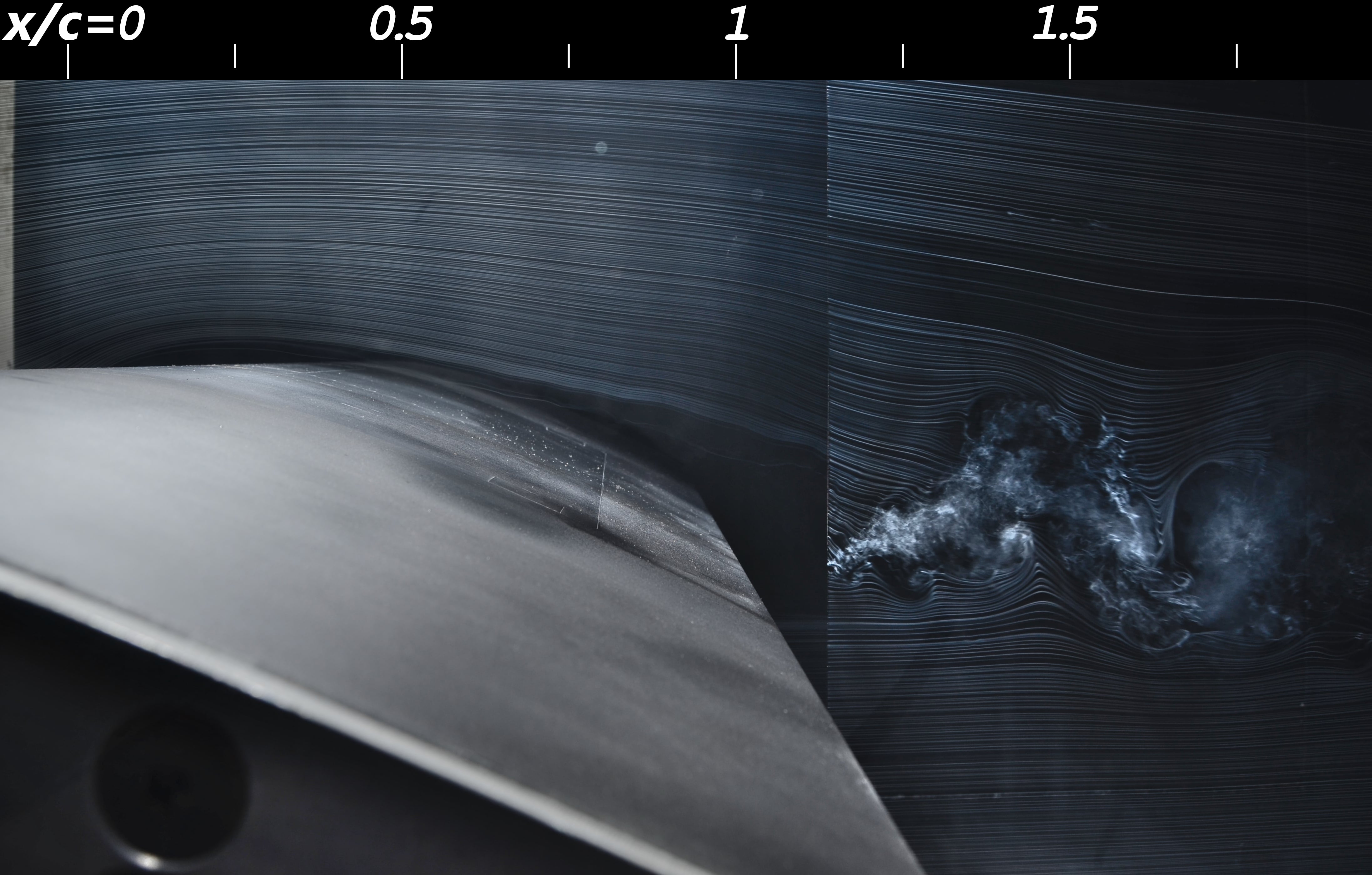}
        \caption{$F^+=11.76$, measured at $z/c=0.17$}
        \label{fig:200Hz5}
    \end{subfigure}
    \begin{subfigure}{0.495\textwidth}
        \includegraphics[width=\linewidth]{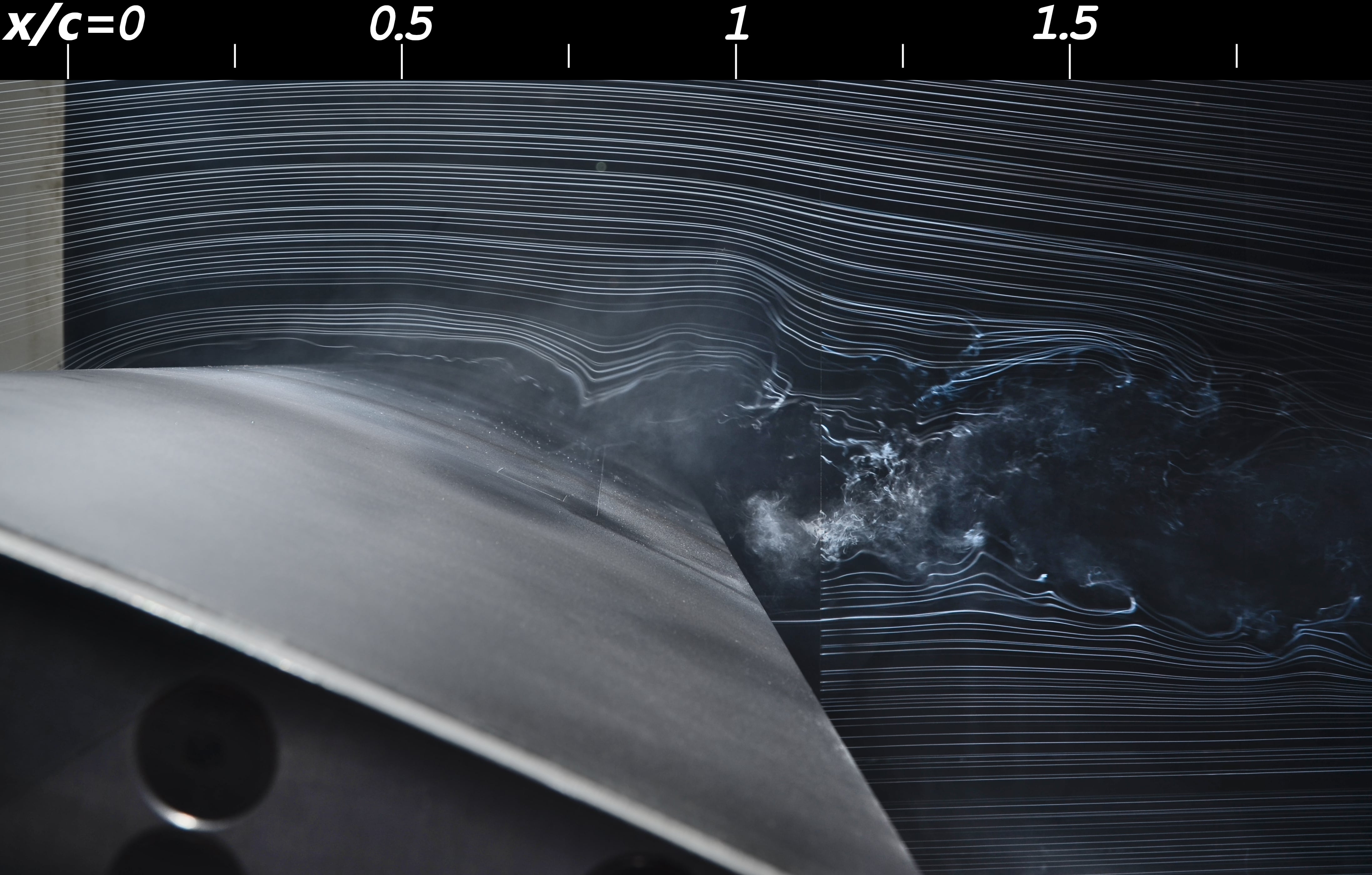}
        \caption{$F^+=1.18$, measured at $z/c=0.33$}
        \label{fig:20Hz10}
    \end{subfigure}
    \begin{subfigure}{0.495\textwidth}
        \includegraphics[width=\linewidth]{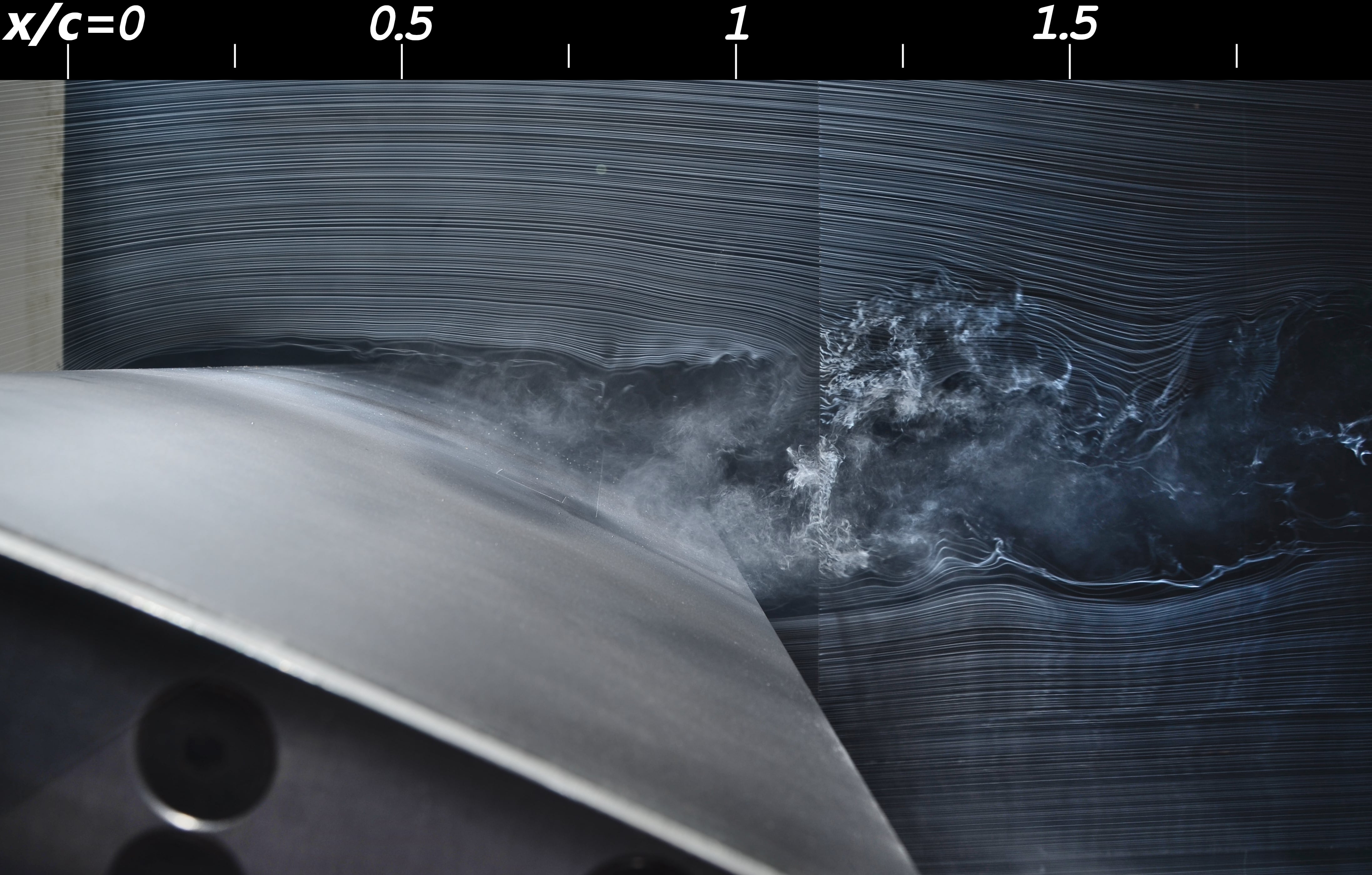}
        \caption{$F^+=11.76$, measured at $z/c=0.33$}
        \label{fig:200Hz10}
    \end{subfigure}
    \caption{Flow with synthetic jet actuation, visualized by both upstream and downstream smoke wires}
    \label{fig:vertical multi-plane}
\end{figure}

Visualizing the controlled flow at midspan (Figures \ref{fig:20Hz0} and \ref{fig:200Hz0}) reveals that in both the low- and high-frequency control cases, the flow reattaches, and the wake size is significantly reduced. The airfoil is no longer stalled in the controlled state, resulting in lift recovery and reduced drag. As seen in Figure \ref{fig:20Hz0}, the low-frequency actuation targeting the wake instability results in a fairly uniform von Kármán vortex street, suggesting a coupling between the shear layer and wake vortices as described by \citet{Glezer2005}. The large lengthscales in this alternating vortex street result in a slightly larger wake, resulting in increased pressure drag. Figure \ref{fig:200Hz0} shows that in the high-frequency control case, the vortex street is seen to be non-uniform, with stronger counter-clockwise vorticity produced from flow under the airfoil. In this case, the shear layer vortices remain uncoupled from the shedding vortices due to their higher frequency and shorter lengthscale. It appears that this discrepancy between the upper and lower layers of the wake enhances the dissipation of the larger, periodic counter-clockwise vortices. The high-frequency control not only results in a smaller wake, but also, reattachment with more steady aerodynamic forces \citep{Xu2023}.

\begin{figure}[!h]
    \centering
    \begin{subfigure}{0.495\textwidth}
        \includegraphics[width=\linewidth]{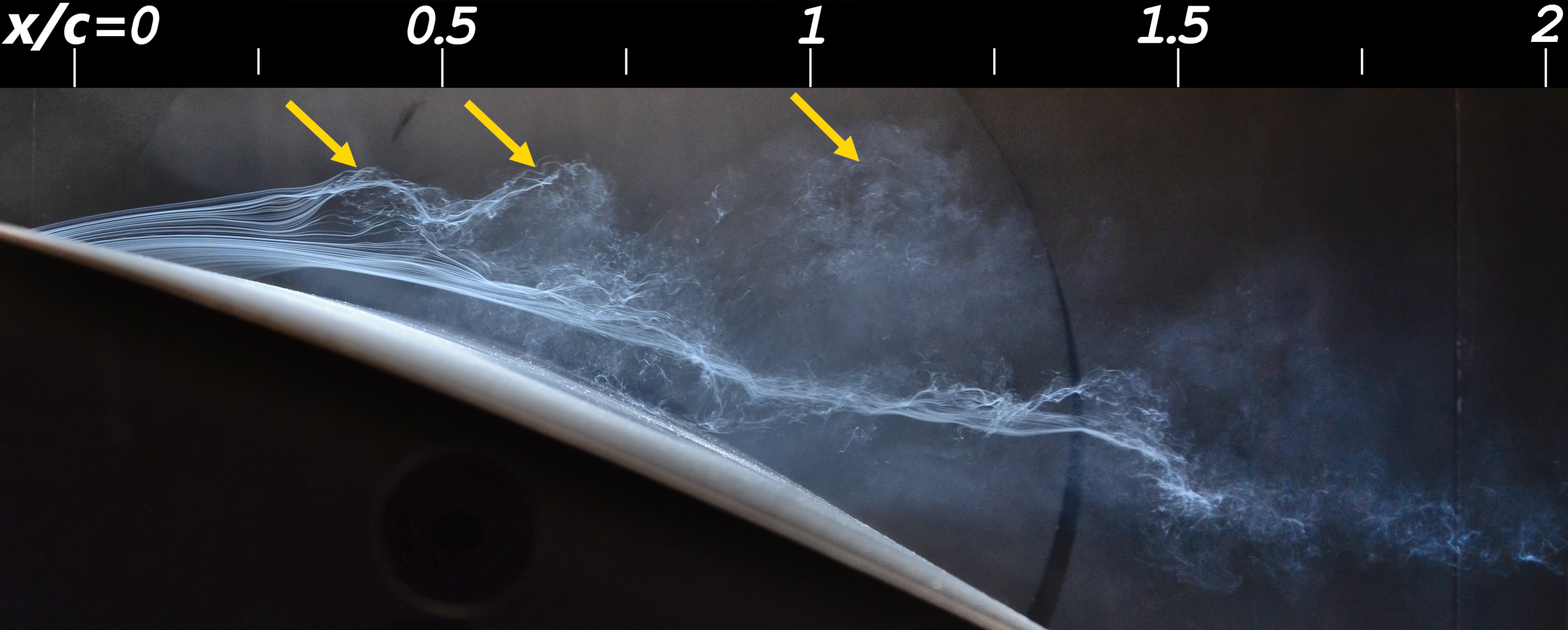}
        \caption{$F^+=1.18$}
        \label{fig:horizontal 20Hz}
    \end{subfigure}
    \begin{subfigure}{0.495\textwidth}
        \includegraphics[width=\linewidth]{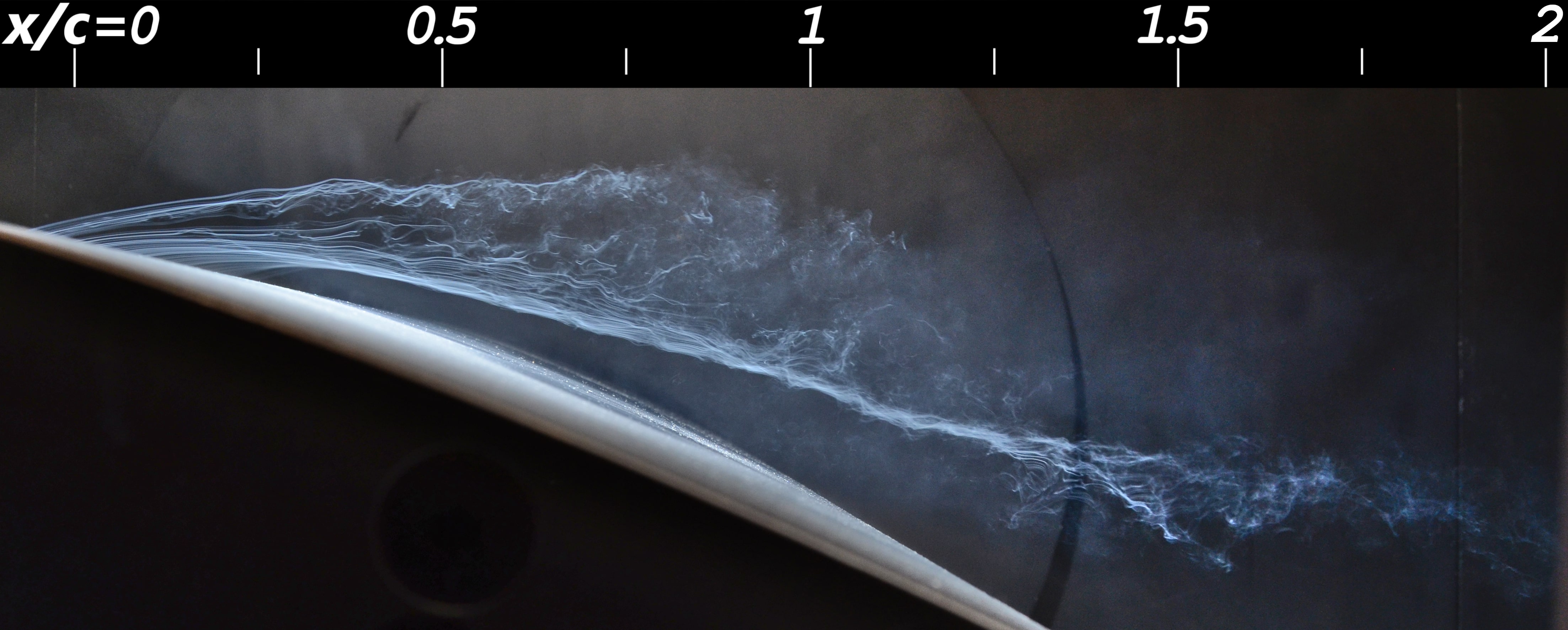}
        \caption{$F^+=11.76$}
        \label{fig:horizontal 200Hz}
    \end{subfigure}
    \caption{Flow with SJ actuation visualized by a horizontal smoke wire and imaged from the side view}
    \label{fig:horizontal}
\end{figure}

The multi-plane flow visualization in Figure \ref{fig:vertical multi-plane} illustrates that the control authority diminishes with increasing distance from midspan. Figures \ref{fig:20Hz5} and \ref{fig:200Hz5} show that the flow measured at $z/c=0.17$ is largely the same as that at midspan; the flow is still attached, and the trailing edge boundary layer thickness measures identically. However, there is a distinguishable difference in the flow dynamics of the wake. For $F^+=11.76$ (Figure \ref{fig:200Hz5}), the vortices shed from the trailing edge are larger in size than at midspan. This increased wake size suggests that the sectional drag coefficient is likely higher at this location, however, the flow is still attached as no recirculation is observed. For both control cases measured at $z/c=0.33$ (Figures \ref{fig:20Hz10} and \ref{fig:200Hz10}), the flow is clearly separated, as indicated by the recirculation at the trailing edge seen from the downstream smoke wire. Additionally, the size of the wake is much larger, and the flow starts to resemble the baseline case. The effect of diminishing control authority away from the midspan is further shown in Figure \ref{fig:horizontal} where the horizontal smoke wire is visualized from the side view. It is observed that while the flow is relatively uniform across the span of the wing in the baseline case (Figure \ref{fig:baselineB}), the controlled flow (Figures \ref{fig:horizontal 20Hz} and \ref{fig:horizontal 200Hz}) possesses spanwise non-uniformity. The flow near the midspan is kept swept to the wing, while the streaklines on the near and far sides of the photograph have a higher trajectory, indicating that the boundary layer is larger at the edge of the array. Furthermore, for the low-frequency control case, a clear pattern indicative of shear layer roll-up is observed from the outermost streaklines (labelled by the three yellow arrows), resembling the baseline flow. These results show that the flow near the edge of the array is highly impacted by the uncontrolled flow that exists past the extent of the array.

For $F^+=1.18$ (Figure \ref{fig:20Hz10}), the streaklines exhibit a sharp downward trajectory at $x/c=0.8$. This phenomenon can be attributed to a large clockwise-rotating vortex above the airfoil, transporting momentum from the freestream into the boundary layer. Nevertheless, reverse flow is observed near the trailing edge, indicating that the boundary layer is not sufficiently energized to overcome the adverse pressure gradient. At the edge of the effective spanwise control region, the shear layer separation is expected to be highly unsteady, oscillating between attached and separated flow, commonly described as shear layer flapping. \citet{Salunkhe2016} observed shear layer flapping over an airfoil when very weak synthetic jet flow control was applied. This suggests that flow at the edge of effective control may have similar dynamics to mildly controlled flow at the symmetry plane. The shear layer flapping is less pronounced for the high-frequency actuation case, likely due to the fact that higher actuation frequencies result in smaller spanwise vortices, and thus more time-invariant control. Conversely, the low-frequency actuation results in larger more spaced apart vortices, giving short bursts of control, and thus a flapping shear layer.

\begin{figure}[t]
    \begin{subfigure}{\textwidth}
        \includegraphics[width=0.495\linewidth]{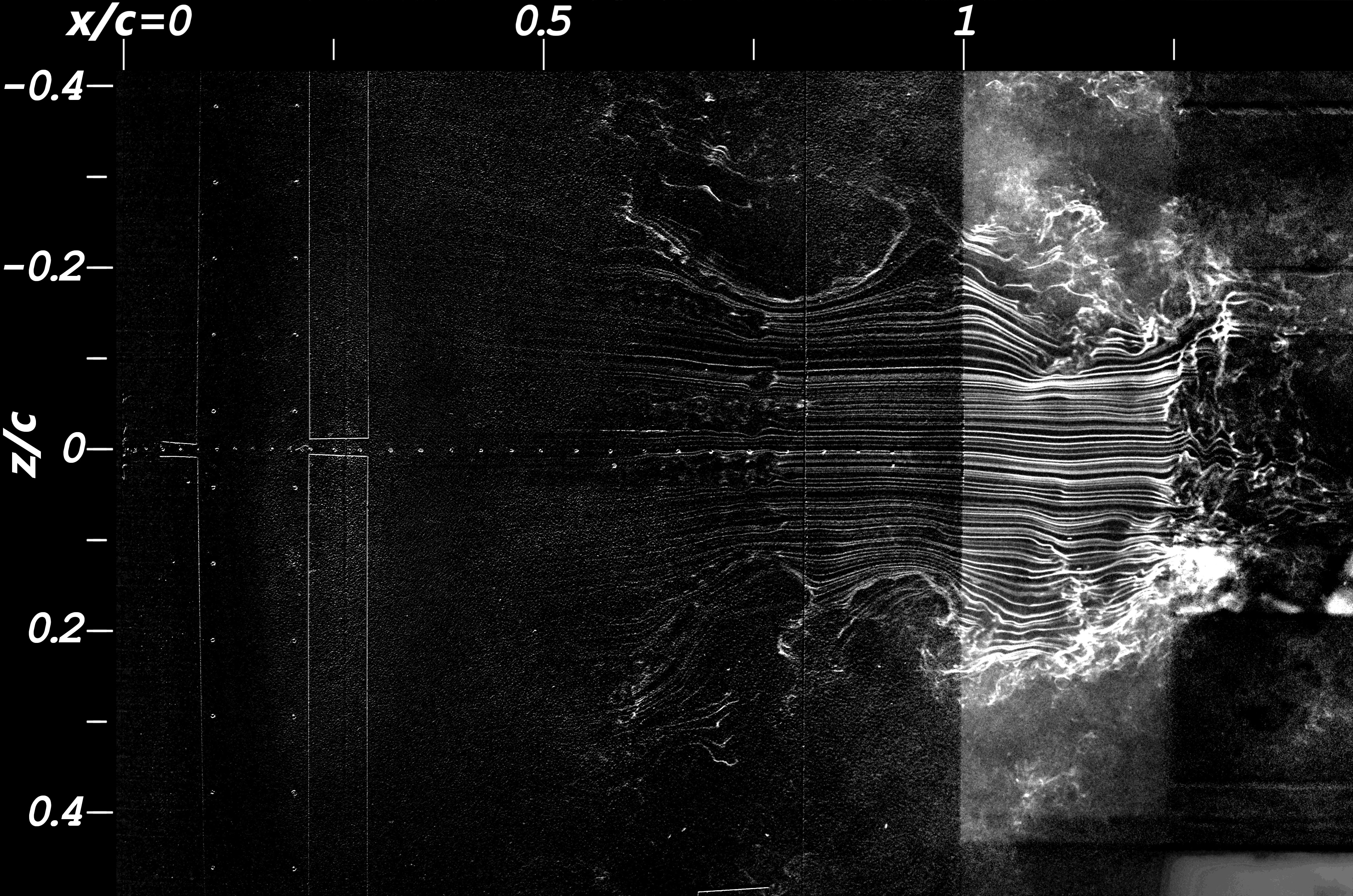}
        \includegraphics[width=0.495\linewidth]{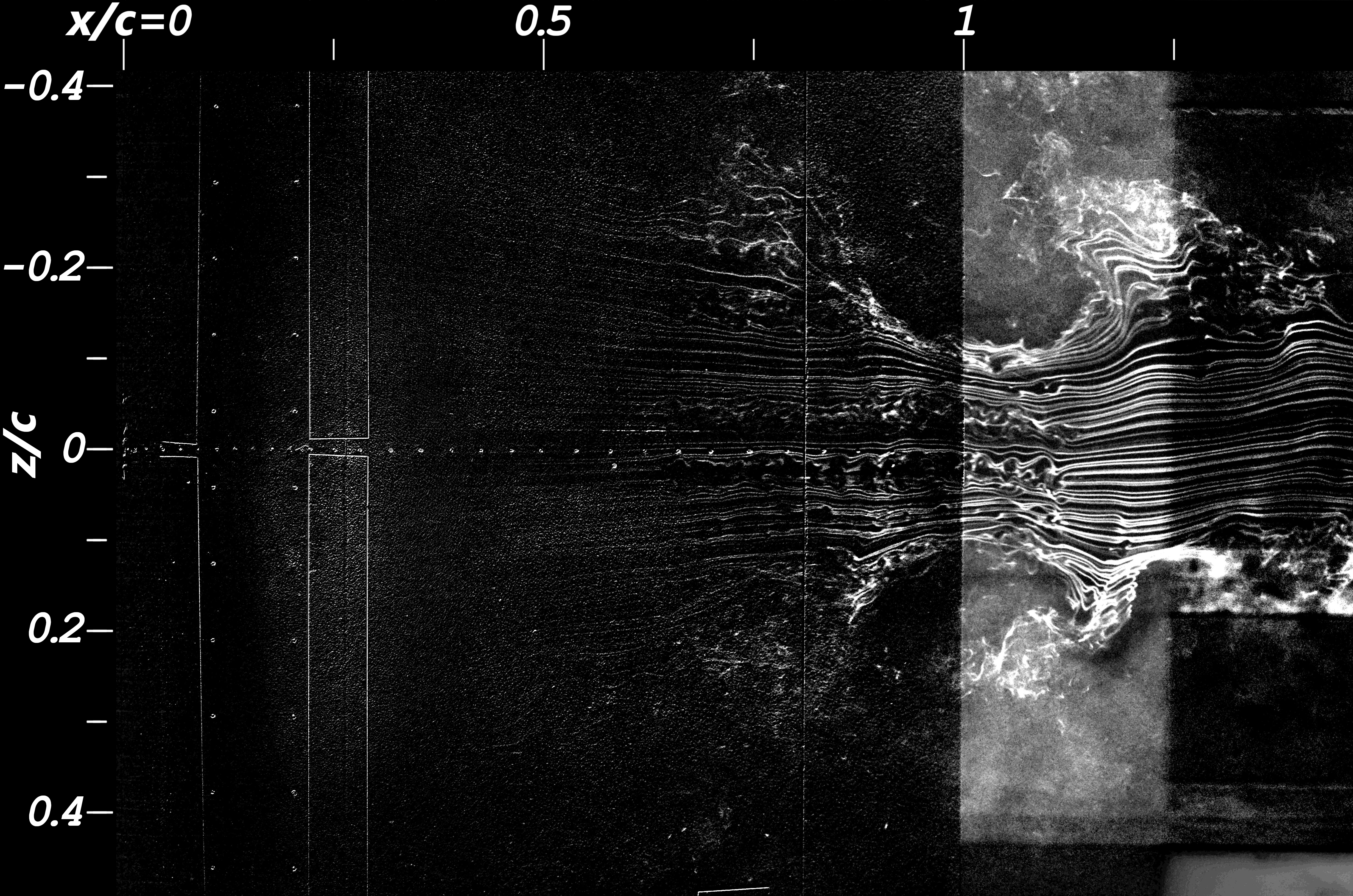}
        \caption{}
        \label{fig:overhead20Hz}
    \end{subfigure}
    \begin{subfigure}{\textwidth}
        \includegraphics[width=0.495\linewidth]{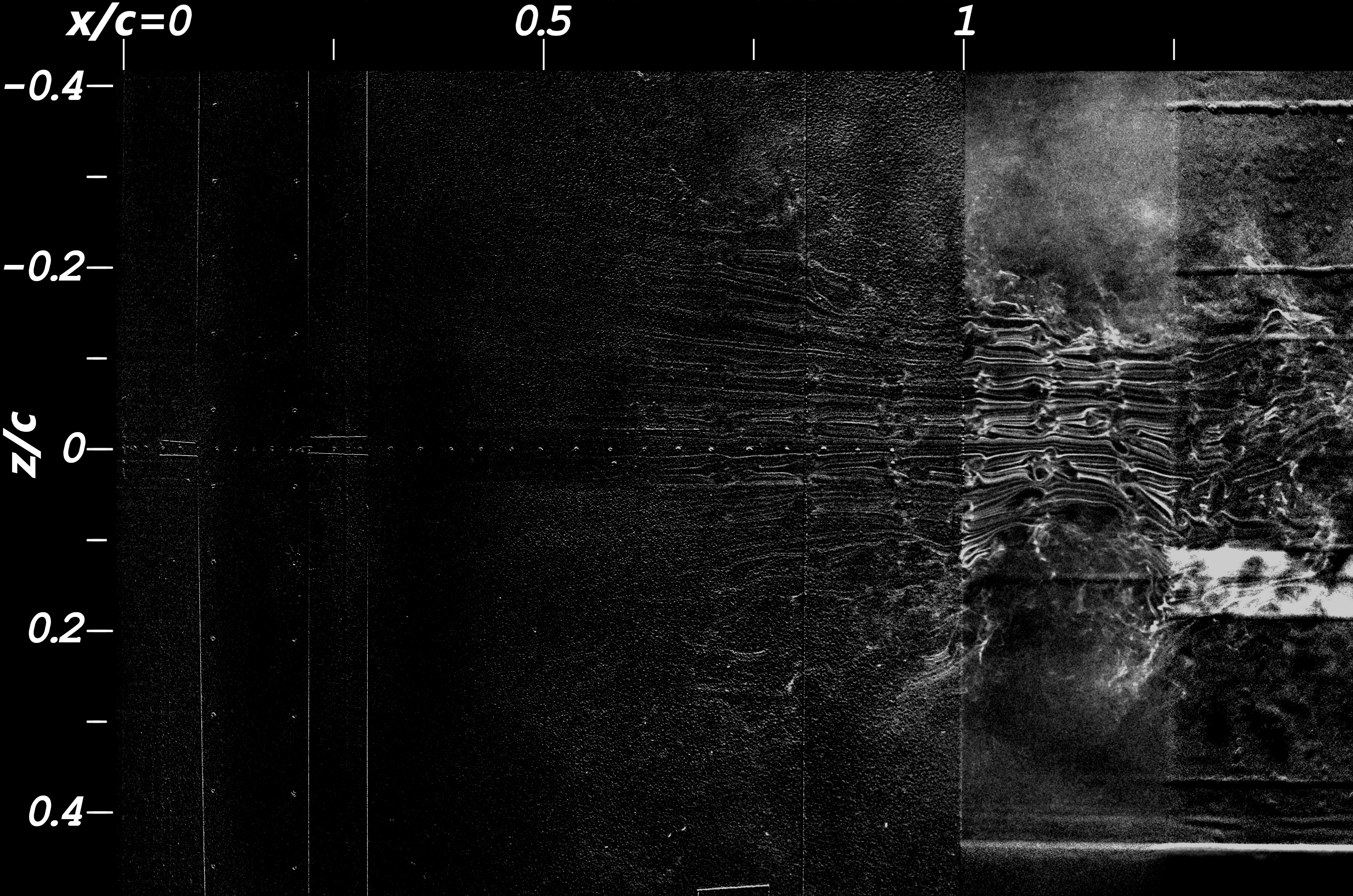}
        \includegraphics[width=0.495\linewidth]{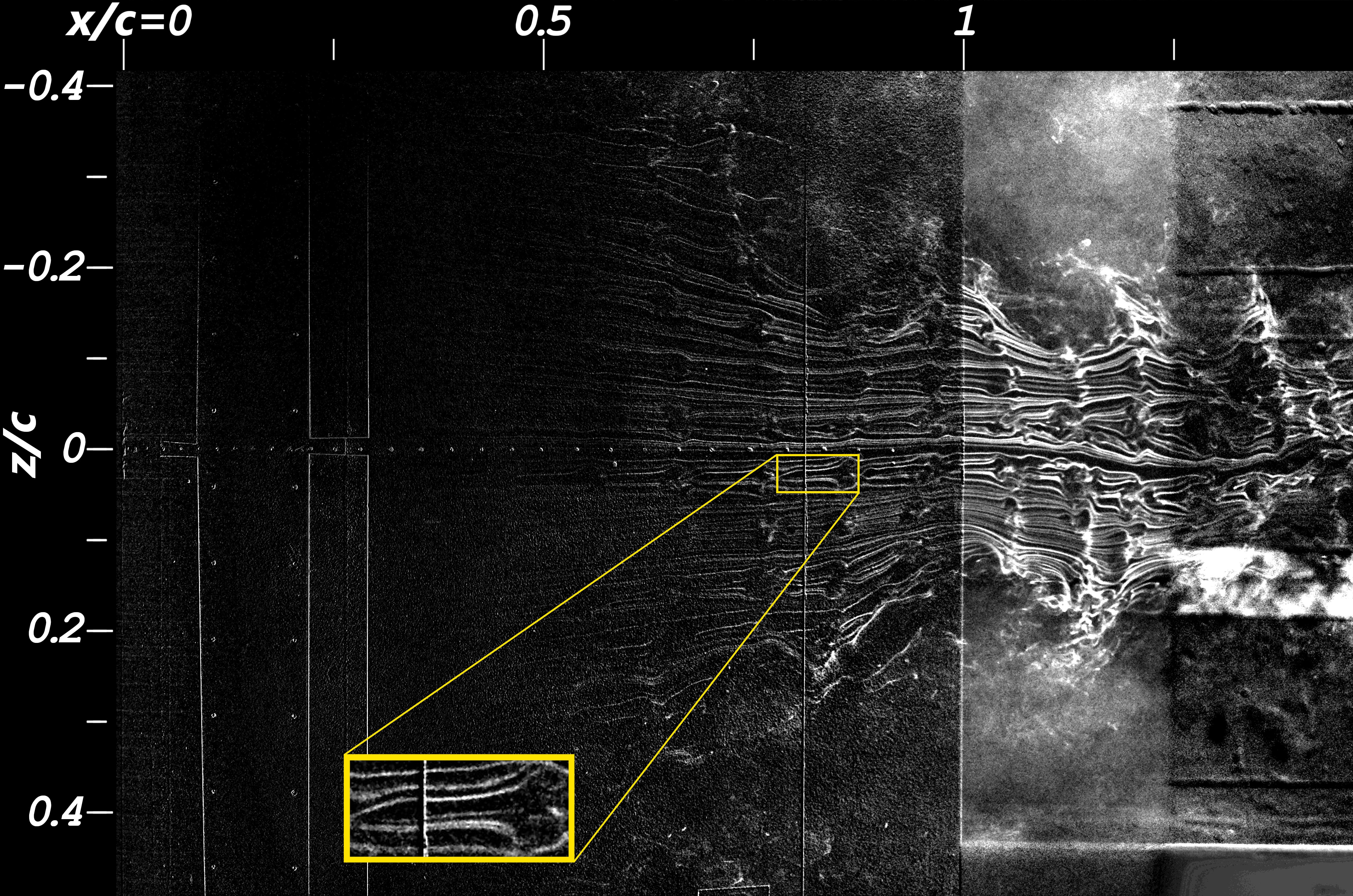}
        \caption{}
        \label{fig:overhead200Hz}
    \end{subfigure}
    \caption{Overhead smoke visualization images with an upstream horizontal wire for a) $F^+=1.18$ and b) $F^+=11.76$. The images on the right were taken 1/6 s after the images on the left}
    \label{fig:overhead}
\end{figure}

The spanwise non-uniformity of the controlled flow suggests that the flow is three-dimensional in nature. The horizontal smoke wire used in conjunction with an overhead camera enables visualization of the flow within the $x$-$z$ plane. The overhead views of the controlled flow in Figure \ref{fig:overhead} reveal a contraction in the flow toward the midspan with both actuation frequencies. This phenomenon is attributed to the separated flow outside the effective control region displacing fluid toward the midspan where faster-moving, and thus lower pressure, fluid exists. Through surface oil flow visualization, \citet{Feero2017a} observed a similar contraction in flow controlled by a slot-style SJA over the same airfoil. Together, these findings show that a flow-control induced contraction exists throughout the entire boundary layer, and is due to the spanwise velocity gradient, i.e. $\partial {u}/ \partial z$.

With actuation at $F^+=1.18$, a sharp contraction is seen in the streaklines followed by a subsequent expansion downstream, forming an hourglass-like pattern (Figure \ref{fig:overhead20Hz}). The left image shows a contraction centered at $x/c\approx0.8$, while in the right image, this contraction is centered at the trailing edge. This observation provides further evidence that low actuation frequencies result in time-variant control, especially when considering the flow away from the midspan. With the same parameters, \citet{Xu2023} observed the presence of large spanwise vortices that roll down the airfoil, with typically one present above the suction surface at a given instant. As discussed in Section \ref{sec:intro}, these roller vortices are said to transport momentum non-uniformly across the chord, resulting in an unsteady flow. Furthermore, the spacing of phase-coherent structures produced by the SJA can be estimated by dividing the freestream velocity by the modulation frequency (i.e., $U_\infty/f_c$), resulting in a spacing of $0.85c$, which is commensurate with the length of the chord. This result confirms that forcing at $F^+=1.18$ results in approximately one spanwise vortex over the airfoil at all times. It should be noted that this spacing is an overestimate since the spanwise vortices of interest are expected to lie within the shear layer and are therefore convected at sub-freestream velocities. It is concluded that the sharp contraction in the flow coincides with the spanwise vortex roller, which energizes the boundary layer locally, thus, resulting in a local contraction of the streaklines.

\newpage

With actuation at $F^+=11.76$, a comparable yet more uniform contraction appears in the streaklines. Additionally, the high-frequency actuation results in distinctive structures which are observed downstream of the SJAs starting at $x/c\approx0.6$, one of which is highlighted by a yellow box and magnified in Figure \ref{fig:overhead200Hz}. Since these patterns manifest in the streaklines that pass directly over the SJA nozzles, it can be inferred that their formation is the direct result of the SJAs. \citet{Xu2023} identified vortices of similar size and frequency in the shear layer using the Q-criterion method on PIV data taken at the midspan. The overhead visualization shows that these structures are most coherent near the midspan where the dissipative effect of the separated outer flow is less prevalent. It is also observed that the structures persist well past the trailing edge, indicating their influence on the wake. Referring to Figure \ref{fig:200Hz0}, the lowest streakline over the airfoil exhibits a growing sinusoidal pattern downstream of $x/c\approx0.6$. Together, the multiple visualization angles reveal the three-dimensional nature of the flow structures. It is hypothesized that the unique pattern in the streaklines is due to the presence of vortex rings that are produced by the SJAs every cycle, and grow in size as they convect downstream. The expansion of the streaklines seen in the magnified region of Figure \ref{fig:overhead200Hz} can be explained by an acceleration of the fluid around the outside of the vortex ring, and the inward curl of the streaklines would be due to the fluid accelerating back toward its core. Vortex rings of the same orientation were observed in a computational study of a synthetic jet in crossflow by \citet{Ho2022}, however their persistence downstream cannot be confirmed due the small computational domain centered around the SJA exit.

\section{Conclusion}
The smoke wire technique was used to visualize the entire flow field over a NACA 0025 profile wing to investigate the effects of synthetic jet flow control. In the baseline flow case the airfoil is stalled, with a large recirculation area and a wide wake. A Kelvin-Helmholtz instability is observed near the leading edge of the airfoil, resulting in roll-up vortices in the shear layer which quickly dissipate as the boundary layer transitions to turbulence. Two actuation frequencies are examined ($F^+=1.18$ and $F^+=11.76$), with particular emphasis on a) spanwise control authority, and b) the attributes of the coherent structures, which are deemed the fundamental mechanism of flow reattachment. While the control is effective at midspan, the flow only reattaches for approximately one-third the length of the array, due to the influence of the outer flow. At an intermediate distance from the midspan, the flow remains attached; however, the diminishing control is indicated by the increase in wake size. At a greater distance from the midspan, flow separation is observed through trailing-edge recirculation. At this spanwise location, the visualization suggests the presence of a flapping shear layer in the low-frequency actuation case, a phenomenon arising from its unsteady reattachment. The overhead visualization of the horizontal smoke wire shows a contraction of the streaklines toward the midspan. For the low-frequency case, this contraction is sharper and is observed at various chordwise locations. In contrast, the high-frequency case results in a more constant and gradual contraction. These observations align with existing literature, supporting the notion that low-frequency forcing leads to unsteady flow dynamics. 

 \bibliography{reference}

\begin{thebibliography}{23}
\providecommand{\natexlab}[1]{#1}
\providecommand{\url}[1]{\texttt{#1}}
\expandafter\ifx\csname urlstyle\endcsname\relax
  \providecommand{\doi}[1]{doi: #1}\else
  \providecommand{\doi}{doi: \begingroup \urlstyle{rm}\Url}\fi

\bibitem[Amitay and Glezer(2002)]{Amitay2002}
Michael Amitay and Ari Glezer.
\newblock Role of actuation frequency in controlled flow reattachment over a stalled airfoil.
\newblock \emph{AIAA Journal}, 40:\penalty0 209--216, 2 2002.
\newblock ISSN 0001-1452.
\newblock \doi{10.2514/2.1662}.

\bibitem[Amitay et~al.(2001)Amitay, Smith, Kibens, Parekh, and Glezer]{Amitay2001}
Michael Amitay, Douglas~R. Smith, Valdis Kibens, David~E. Parekh, and Ari Glezer.
\newblock Aerodynamic flow control over an unconventional airfoil using synthetic jet actuators.
\newblock \emph{AIAA Journal}, 39:\penalty0 361--370, 3 2001.
\newblock ISSN 0001-1452.
\newblock \doi{10.2514/2.1323}.

\bibitem[Andino et~al.(2015)Andino, Lin, Washburn, Whalen, Graff, and Wygnanski]{Andino2015}
Marlyn~Y. Andino, John~C. Lin, Anthony~E. Washburn, Edward~A. Whalen, Emilio~C. Graff, and Israel~J. Wygnanski.
\newblock Flow separation control on a full-scale vertical tail model using sweeping jet actuators.
\newblock American Institute of Aeronautics and Astronautics, 1 2015.
\newblock ISBN 978-1-62410-343-8.
\newblock \doi{10.2514/6.2015-0785}.

\bibitem[Balzer and Fasel(2010)]{Balzer2010}
Wolfgang Balzer and Hermann Fasel.
\newblock Direct numerical simulation of laminar boundary-layer separation and separation control on the suction side of an airfoil at low reynolds number conditions.
\newblock American Institute of Aeronautics and Astronautics, 6 2010.
\newblock ISBN 978-1-60086-956-3.
\newblock \doi{10.2514/6.2010-4866}.

\bibitem[Batill and Mueller(1981)]{Batill1981}
Stephen~M. Batill and Thomas~J. Mueller.
\newblock Visualization of transition in the flow over an airfoil using the smoke-wire technique.
\newblock \emph{AIAA Journal}, 19:\penalty0 340--345, 3 1981.
\newblock ISSN 0001-1452.
\newblock \doi{10.2514/3.50953}.

\bibitem[Cicca and Iuso(2007)]{DiCicca2007}
Gaetano Maria~Di Cicca and Gaetano Iuso.
\newblock On the near field of an axisymmetric synthetic jet.
\newblock \emph{Fluid Dynamics Research}, 39:\penalty0 673--693, 9 2007.
\newblock ISSN 0169-5983.
\newblock \doi{10.1016/j.fluiddyn.2007.03.002}.

\bibitem[Feero et~al.(2015)Feero, Goodfellow, Lavoie, and Sullivan]{Feero2015}
Mark~A. Feero, Sebastian~D. Goodfellow, Philippe Lavoie, and Pierre~E. Sullivan.
\newblock Flow reattachment using synthetic jet actuation on a low-reynolds-number airfoil.
\newblock \emph{AIAA Journal}, 53:\penalty0 2005--2014, 7 2015.
\newblock ISSN 0001-1452.
\newblock \doi{10.2514/1.J053605}.

\bibitem[Feero et~al.(2017a)Feero, Lavoie, and Sullivan]{Feero2017a}
Mark~A. Feero, Philippe Lavoie, and Pierre~E. Sullivan.
\newblock Three-dimensional span effects of high-aspect ratio synthetic jet forcing for separation control on a low reynolds number airfoil.
\newblock \emph{Journal of Visualization}, 20, 2017a.
\newblock ISSN 18758975.
\newblock \doi{10.1007/s12650-016-0365-7}.

\bibitem[Feero et~al.(2017b)Feero, Lavoie, and Sullivan]{Feero2017b}
Mark~A. Feero, Philippe Lavoie, and Pierre~E. Sullivan.
\newblock Influence of synthetic jet location on active control of an airfoil at low reynolds number.
\newblock \emph{Experiments in Fluids}, 58, 2017b.
\newblock ISSN 07234864.
\newblock \doi{10.1007/s00348-017-2387-x}.

\bibitem[Glezer et~al.(2005)Glezer, Amitay, and Honohan]{Glezer2005}
Ari Glezer, Michael Amitay, and Andrew~M. Honohan.
\newblock Aspects of low- and high-frequency actuation for aerodynamic flow control.
\newblock \emph{AIAA Journal}, 43:\penalty0 1501--1511, 7 2005.
\newblock ISSN 0001-1452.
\newblock \doi{10.2514/1.7411}.

\bibitem[Greenblatt and Wygnanski(2000)]{Greenblatt2000}
David Greenblatt and Israel~J. Wygnanski.
\newblock The control of flow separation by periodic excitation.
\newblock \emph{Progress in Aerospace Sciences}, 36:\penalty0 487--545, 10 2000.
\newblock ISSN 03760421.
\newblock \doi{10.1016/S0376-0421(00)00008-7}.

\bibitem[Ho et~al.(2022)Ho, Essel, and Sullivan]{Ho2022}
Haonan~H. Ho, Ebenezer~E. Essel, and Pierre~E. Sullivan.
\newblock The interactions of a circular synthetic jet with a turbulent crossflow.
\newblock \emph{Physics of Fluids}, 34, 7 2022.
\newblock ISSN 1070-6631.
\newblock \doi{10.1063/5.0099533}.

\bibitem[Kirk and Yarusevych(2017)]{Kirk2017}
Thomas~M. Kirk and Serhiy Yarusevych.
\newblock Vortex shedding within laminar separation bubbles forming over an airfoil.
\newblock \emph{Experiments in Fluids}, 58:\penalty0 43, 5 2017.
\newblock ISSN 0723-4864.
\newblock \doi{10.1007/s00348-017-2308-z}.

\bibitem[Rice et~al.(2018)Rice, Taylor, and Amitay]{Rice2018}
Thomas~T. Rice, Keith Taylor, and Michael Amitay.
\newblock Quantification of the s817 airfoil aerodynamic properties and their control using synthetic jet actuators.
\newblock \emph{Wind Energy}, 21:\penalty0 823--836, 10 2018.
\newblock ISSN 10954244.
\newblock \doi{10.1002/we.2197}.

\bibitem[Salunkhe et~al.(2016)Salunkhe, Tang, Zheng, and Wu]{Salunkhe2016}
Pramod Salunkhe, Hui Tang, Yingying Zheng, and Yanhua Wu.
\newblock Piv measurement of mildly controlled flow over a straight-wing model.
\newblock \emph{International Journal of Heat and Fluid Flow}, 62:\penalty0 552--559, 12 2016.
\newblock ISSN 0142727X.
\newblock \doi{10.1016/j.ijheatfluidflow.2016.08.004}.

\bibitem[Seifert and Pack(1999)]{Seifert1999}
A.~Seifert and L.~G. Pack.
\newblock Oscillatory control of separation at high reynolds numbers.
\newblock \emph{AIAA Journal}, 37:\penalty0 1062--1071, 9 1999.
\newblock ISSN 0001-1452.
\newblock \doi{10.2514/2.834}.

\bibitem[Smith and Swift(2003)]{Smith2003}
B.~L. Smith and G.~W. Swift.
\newblock A comparison between synthetic jets and continuous jets.
\newblock \emph{Experiments in Fluids}, 34:\penalty0 467--472, 4 2003.
\newblock ISSN 0723-4864.
\newblock \doi{10.1007/s00348-002-0577-6}.

\bibitem[Tang et~al.(2014)Tang, Salunkhe, Zheng, Du, and Wu]{Tang2014}
Hui Tang, Pramod Salunkhe, Yingying Zheng, Jiaxing Du, and Yanhua Wu.
\newblock On the use of synthetic jet actuator arrays for active flow separation control.
\newblock \emph{Experimental Thermal and Fluid Science}, 57:\penalty0 1--10, 9 2014.
\newblock ISSN 08941777.
\newblock \doi{10.1016/j.expthermflusci.2014.03.015}.

\bibitem[Tousi et~al.(2021)Tousi, Coma, Bergadà, Pons-Prats, Mellibovsky, and Bugeda]{Tousi2021}
N.M. Tousi, M.~Coma, J.M. Bergadà, J.~Pons-Prats, F.~Mellibovsky, and G.~Bugeda.
\newblock Active flow control optimisation on sd7003 airfoil at pre and post-stall angles of attack using synthetic jets.
\newblock \emph{Applied Mathematical Modelling}, 98:\penalty0 435--464, 10 2021.
\newblock ISSN 0307904X.
\newblock \doi{10.1016/j.apm.2021.05.016}.

\bibitem[Toyoda and Hiramoto(2009)]{Toyoda2009}
Kuniaki Toyoda and Riho Hiramoto.
\newblock Manipulation of vortex rings for flow control.
\newblock \emph{Fluid Dynamics Research}, 41:\penalty0 051402, 10 2009.
\newblock ISSN 0169-5983.
\newblock \doi{10.1088/0169-5983/41/5/051402}.

\bibitem[Xu et~al.(2023)Xu, Lavoie, and Sullivan]{Xu2023}
Kecheng Xu, Philippe Lavoie, and Pierre Sullivan.
\newblock Flow reattachment on a naca 0025 airfoil using an array of microblowers.
\newblock \emph{AIAA Journal}, pages 1--10, 3 2023.
\newblock ISSN 0001-1452.
\newblock \doi{10.2514/1.J062512}.

\bibitem[Yang et~al.(2022)Yang, Ekmekci, and Sullivan]{Yang2022}
Eric Yang, Alis Ekmekci, and Pierre~E. Sullivan.
\newblock Phase evolution of flow controlled by synthetic jets over naca 0025 airfoil.
\newblock \emph{Journal of Visualization}, 25:\penalty0 751--765, 8 2022.
\newblock ISSN 1343-8875.
\newblock \doi{10.1007/s12650-021-00824-5}.

\bibitem[Yarusevych et~al.(2008)Yarusevych, Kawall, and Sullivan]{Yarusevych2008}
Serhiy Yarusevych, John~G. Kawall, and Pierre~E. Sullivan.
\newblock Separated-shear-layer development on an airfoil at low reynolds numbers.
\newblock \emph{AIAA Journal}, 46:\penalty0 3060--3069, 2008.
\newblock ISSN 00011452.
\newblock \doi{10.2514/1.36620}.

\end{thebibliography}
\end{document}